\documentclass[journal]{IEEEtran}
\usepackage[colorlinks]{hyperref}
\usepackage{cite}
\usepackage{mathrsfs}
\usepackage{bbding}
\usepackage{amsthm}
\usepackage{caption}
\usepackage{etoolbox} \makeatletter \patchcmd{\@makecaption} {\scshape} {} {} {} \makeatother
\usepackage[cmex10]{amsmath}
\usepackage{float}
\usepackage{mathtools}
 \usepackage{siunitx}
\usepackage{array}
\usepackage{eqparbox}
\usepackage{bm}

\usepackage[table,dvipsnames]{xcolor}
\usepackage{multicol,booktabs,tabularx}

\usepackage{cases}
\usepackage{algorithm}
\usepackage{paralist}
\usepackage{algpseudocode}

\usepackage[utf8]{inputenc}
\usepackage[english]{babel}
\usepackage{graphicx}
\usepackage{subfigure}
\usepackage{epstopdf}
\usepackage{epsfig}
\usepackage{amssymb}
\usepackage{array}
\usepackage{multirow}

\usepackage{caption}
\usepackage[numbers,sort&compress]{natbib} % 按引用顺序排序
\begin{document}

\title{Generalizable and Computational Efficient Channel Extrapolation for 6G: A Configurable AI-Driven Framework Built from a Modular Perspective}
% TODO 重点Time-Aware
\author{Yuan Gao$^1$\Envelope,
Xinyi Wu$^1$, 
Jiang Jun$^1$,
Yi Yu$^2$,
Yanliang Jin$^1$,
%Jianbo Du,~\IEEEmembership{Member,~IEEE},
Shunqing Zhang$^1$,
%Weijie Yuan,
%Cunhua Pan,~\IEEEmembership{Senior Member,~IEEE},
%Xiaoli Chu,~\IEEEmembership{Senior Member,~IEEE},\\
%Honggang Zhang,~\IEEEmembership{Fellow,~IEEE},
%Merouane Debbah,~\textit{Fellow, IEEE},
Zhu Han$^3$,
and Shugong Xu$^4$,

%\thanks{Yuan Gao, Xinyi Wu and Shunqing Zhang are with the School of Communication and Information Engineering, Shanghai University, China, email: gaoyuansie@shu.edu.cn, wu\_xinyi0312@shu.edu.cn and shunqing@shu.edu.cn.}
\thanks{$^1$School of Communication and Information Engineering, Shanghai University, China. $^2$Division of Engineering, New York University Abu Dhabi, Abu Dhabi, United Arab Emirates. $^3$Department of Electrical and Computer Engineering at the University of Houston, Houston, USA. $^4$School of Advanced Technology, Xi'an Jiaotong-Liverpool University, Suzhou, China.}
\thanks{\Envelope gaoyuansie@shu.edu.cn.}
}

\maketitle

% As a general rule, do not put math, special symbols or citations in the abstract or keywords.
\begin{abstract}
Acquiring channel state information (CSI) with manageable overhead has been essential to provide high-performance communication services, which is extremely challenging in the emerging sixth generation (6G) mobile network. Channel extrapolation has been proposed to infer complete CSI using a small portion of known CSI, its performance can be dramatically enhanced by artificial intelligence (AI). However, AI-driven channel extrapolation suffers from poor generalization across scenarios and high computational complexity, which is common in the broad research of AI and large language models. Inspired by the modular function of human brain, we propose a configurable AI-driven framework to achieve generalizable and computational efficient channel extrapolation from a modular perspective. We propose a three-stage framework, consisting of experts emergent, experts construction and experts selection. This framework assumes that CSI correlations can be captured by a small number of specialized functional modules (experts) that are activated differently across scenarios. Such modularity emerges in the experts emergent stage via pre-training using CSI data covering comprehensive scenarios. The neurons with similar weight-space patterns are grouped as experts in the experts construction stage. A lightweight gating function is added to control the routing of experts and is fine-tuned for each scenario in the experts selection stage. Simulation results demonstrate that the proposed three-stage framework reduce the channel extrapolation error and computational complexities dramatically by $1.1-19.1$ db and $38$ \%, respectively. In addition, attributed to the proposed experts emergent and section modules, the proposed framework outperforms its counterpart mix-of-expert model dramatically in terms of channel extrapolation performance. 
\end{abstract}


% \fi
\IEEEpeerreviewmaketitle

\section*{Introduction}
With the evolution of the mobile networks, the sixth generation (6G) is expected to be commercialized in 2030, providing unprecedented communication, positioning and sensing services \cite{gao2026sidelink,xu2025enhanced,gao2025stochastic,wang2023road}. To achieve the above vision, sophisticated schemes for traffic prediction \cite{wang2025hyper}, radio resource allocation \cite{hu2024performance,gao2024performance}, interference management \cite{gao2023fair,hu2020fairness,gao2019licensed,chen2016coexistence}, beam management \cite{jin2026generalizable} etc., are required. The effectiveness of the above schemes are highly-related to the knowledge of the channel state information (CSI) \cite{gao2023matching}, acquiring accurate CSI with manageable overhead has been essential for mobile networks \cite{gao2026multiass,zhang2023ai}. Channel extrapolation has emerged as a promising solution to effectively acquire CSI with low overhead by inferring CSI of interest using limited number of known CSI \cite{gao2026effective}.

Channel extrapolation schemes can be classified into model-based and artificial intelligence (AI)-driven approaches \cite{gao2025enabling}. Model-based approaches acquire CSI of interest using limited number of known CSI based on explicit mathematical assumption/modeling of the wireless channel. For example, autoregressive (AR) processes-based approaches assume the CSI of interest as a linear combination of its historical CSI sequence plus noise \cite{liu2022massive,wang2024two,zhang2023basis}. Parametric wireless channel model-based approaches models the wireless channel with respect to key factors including complex amplitudes, delays, and Doppler shifts \cite{peng2022novel, xu2022sparse,xiao2023nonparametric}. Channel extrapolation is achieved by estimating the above factors and calculate the CSI of interest accordingly. However, the model-based schemes suffers from dramatically performance degradation if the mathematical assumption/modeling of the wireless channel mismatch with the actual signal propagation characteristics \cite{gao2026csiextra}. This is quite normal for 6G, which is expected to cover a wide range of scenarios, ranging from space, sky to terrestrial. 

AI-driven approaches avoid explicit mathematical assumption/modeling on the wireless channel by using sophisticated AI models to learn the correlation of CSI and infer the CSI of interest \cite{zhang2023ai,gao2026csiextra}. Various AI models have been investigated for channel extrapolation, from multilayer perceptron (MLP) \cite{kim2020massive,chen2024channel}, recurrent neural network (RNN)-series models \cite{stenhammar2024comparison,jiang2019recurrent,peng2020lstm}, convolutional neural network (CNN) \cite{zhang2022predicting}, Transformer \cite{fan2024e2enet,jin2025linformer}, etc., and have improved the performance of channel extrapolation \cite{SSnet2025gao}. Scaling law has inspired the research of AI-enabled wireless communication, and large AI models trained using massive wireless dataset have been proposed. The efforts including using the large language models (LLMs) in channel extrapolation \cite{fan2025csi,lin2025generative,catak2025bert4mimo}, or developing dedicated foundation model for channel-related tasks \cite{sheng2025wireless,jiang2025mimo}.

Despite the significant improvement of AI-driven channel extrapolation, the current research exhibits several non-trivial {\em limitations}. 

 6G is expected to support a wide range of scenarios, which differentiate each other in terms of signal propagation characteristics dramatically. The existing research mainly focus on improving the channel extrapolation accuracy, but pay much less attention to the generalization performance in unseen scenarios, which has been considered as a key performance indicator for AI-enabled wireless communication by industry \cite{gao2025enabling,3gpp38843}. 

 Either LLM-based or foundation model-based channel extrapolation approaches excel over the conventional AI models, such as RNN or CNN, at the expense of surging computational complexity. Transformer has been the core module for state-of-the-art channel extrapolation approaches, although a certain level of sparsification has been considered for the model design, for example sparse attention mechanisms and mixed of experts (MoE), the computational complexity is reduced at the expense of extrapolation performance\cite{SSnet2025gao}.  

    %\item \textbf{The lack of explainability}: AI-driven approaches work as black boxes, which makes is almost impossible to understand and interpret the fundamental reasons if the channel extrapolation performance is poor, not to mention figuring out solutions to improve its performance systemically. 
    %\item \textbf{Redundant neurons activation and waste of computation resource}: Transformer has been the core module for state-of-the-art channel extrapolation approaches, although a certain level of sparsification has been considered for the model design, such characteristics has not yet be considered for model design systemically. 

The above limitations are not unique to channel extrapolation/wireless communication but are widespread in broader AI/LLM research \cite{lake2017building,matarazzo2025survey}. Compared with the `artificial' intelligence, natural intelligence, i.e., human brain, is far more sophisticated \cite{gebicke2023computational}. On one hand, brain shows dramatic generalization ability. For example, a 3–5-year-old child, after seeing just 1–2 photos or real-life examples of a giraffe, can immediately recognize it as `this is a giraffe' under any viewing angle, lighting condition, cartoon style, or partial occlusion, and further generalize to novel scenarios such as `a giraffe in the zoo' or `a giraffe' eating leaves \cite{vinyals2016matching}. On the other, the human brain achieves extraordinary computational efficiency by performing complex tasks with minimal overall computational resource usage, it rarely activates the entire brain simultaneously, but instead sparsely engages only a small percentage of neurons and dynamically configures combinations of specialized functional modules according to the task demands \cite{attwell2001energy}. One of the main reasons for the outstanding performance of brain is that the brain operates in a modular manner, and the research on AI and LLM began to study and analyze LLMs from a modular perspective \cite{pfeiffer2023modular,ding2023parameter,zhang2023emergent}. Human brain comprises functionally specialized modules, such as the Broca’s area for speech production, Wernicke's area for language comprehension, visual cortex for visual inputs processing, and auditory cortex for speech processing \cite{marshall2013discoveries}. These modules collaborate to generate integrated cognition and behavior for complicated tasks, and only relevant modules will be activated to reduce the computation consumption. To this end, there is a growing tendency to decompose LLMs into different functional modules, allowing for inference with part of modules and dynamic assembly of modules to tackle complex tasks, thereby improving the computational efficiency and generalization \cite{zador2022toward,xiao2024configurable}. 

To this end, inspired by both the functionality of brain and the latest research advancement of the AI/LLM in a modular perspective, we propose a novel configurable AI-driven channel extrapolation framework to cope with the computational efficiency and generalization of channel extrapolation task. The main contributions of this paper are summarized as follows:

 We developed a configurable AI-driven channel extrapolation framework from a modular perspective. The framework treats CSI correlation as being captured by a small set of specialized functional experts that are selectively activated according to the input scenario. To the best of our knowledge, this is the first AI-driven model for wireless communications constructed in a modular manner via expert emergence, clustering and selection.

    The proposed configurable AI-driven channel extrapolation framework contains expert-emerging, expert-construction and expert-selection stage. In the expert-emerging stage, we train a masked auto-encoder (MAE) \cite{he2022masked} using a mixed dataset comprising CSI data in indoor LoS and outdoor NLoS scenarios, to make the modular experts for wireless signal emerge. In the expert-construction stage, we cluster the neurons according to weight-space similarity in the feedforward network (FFN) into dedicated experts. In the expert-selection stage, we adopt MLP as the gating module and fine-tune the remaining framework apart from the experts. After the above three-stage framework, we obtain the same `MoE' model proposed in \cite{SSnet2025gao} in a fundamental different way from a modular perspective.

    Simulation results demonstrate that the channel extrapolation accuracy and computational complexities could be dramatically improved by the proposed three-stage framework. In addition, the proposed framework outperforms its counterpart MoE model, with comparable computational complexities, in terms of channel extrapolation error by $0.8-16.2$ dB and $1.1-18.3$ dB in indoor scenarios and outdoor scenarios, respectively. This validate that the proposed experts emergent and section modules effectively capture the channel characteristics among various propagation environments. Further, we validate that the cosine similarity of the experts' activation probability between scenarios echo with the intuitive understanding of the similarities between scenarios.

%The remaining paper is organized as follows. We formulate the problem of channel extrapolation in Section \ref{Sec:2}. In Section \ref{Sec:3}, we elaborate the proposed channel extrapolation framework, particularly the three-stage framework consisting of experts emergent, experts construction and experts selection. Extensive simulation results with in-depth analysis are presented in Section \ref{Sec:4}. Finally, Section \ref{Sec:5} ends this paper with conclusions.

\begin{figure*}[htbp]
  \centering
  \includegraphics[width=0.83\linewidth]{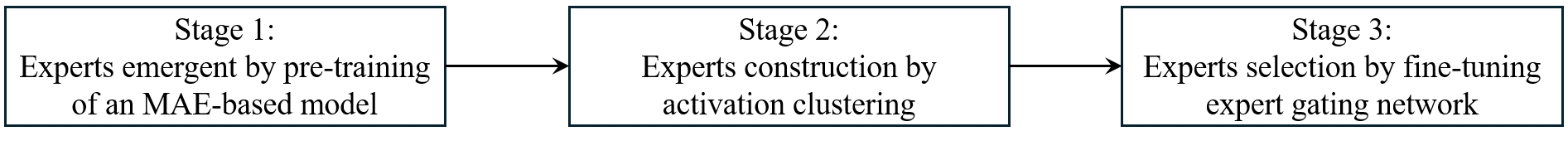} 
  \caption{Proposed three-stage framework. Architecture of the proposed configurable channel extrapolation framework, which pre-trains a model using masked auto-encoder (MAE), then reorganizes the encoder feed-forward network (FFN) layers into specialized expert modules, and finally fine-tunes the adapted model separately across diverse propagation scenarios.}
  \label{fig:whole_arch}
\end{figure*}
\section*{Methods}
\subsection*{Problem formulation of channel extrapolation}
\label{Sec:2}
We consider a TDD MIMO system, where the base station (BS) and user equipment (UE) are equipped with $N_\text{T}$ and $N_\text{R}$ antennas, respectively. Without loss of generality, we assume perfect DL/UL reciprocity and investigate the DL channel between the BS and UE, which is denoted by the channel matrix $\textbf{H} \in \mathbb{C}^{N_\text{R} \times N_\text{T}}$ as:

\begin{equation}
  \textbf{H} = \left(
    \begin{array}{ccc}
      h_{11} & \cdots & h_{1N_\text{T}}\\
      \vdots & \ddots & \vdots\\
      h_{N_\text{R}1} & \cdots & h_{N_\text{R}N_\text{T}}\\
    \end{array}
  \right),
\end{equation}
where $h_{n_\text{R}n_\text{T}}$ is the CSI between the $n_\text{R}$-th antenna at the UE and the $n_\text{T}$-th antenna at the BS.

To capture the wireless propagation characteristics, such as the prorogation attenuation, multi-path effect, etc, we consider the following channel model:
\begin{equation}
h_{n_\text{R}n_\text{T}}=\sum_{l_{n_\text{R}n_\text{T}}=1}^{L_{n_\text{R}n_\text{T}}}\alpha_{l_{n_\text{R}n_\text{T}}} e^{j \phi_{l_{n_\text{R}n_\text{T}}}} \mathbf{A}_\text{R}\mathbf{A}_\text{T}^H,
\end{equation}
where $L_{n_\text{R}n_\text{T}}$ is the number of propagation paths between the $m$-th antenna at the BS and the $k$-th antenna at UE. $\alpha_{l_{n_\text{R}n_\text{T}}}$ and $\phi_{l_{n_\text{R}n_\text{T}}}$ are the amplitude and phase of the propagation path $L_{n_\text{R}n_\text{T}}$, respectively. $\mathbf{A}_\text{R}$ is the receive array response vector with respect to the azimuth and elevation angles of arrival (AoA) $\theta_{l_{n_\text{R}n_\text{T}}}^r$ and $\phi_{l_{n_\text{R}n_\text{T}}}^r$. $\mathbf{A}_\text{T}$ is the transmit array response vector with respect to the azimuth and elevation angles of departure (AoD) $\theta_{l_{n_\text{R}n_\text{T}}}^t$ and $\phi_{l_{n_\text{R}n_\text{T}}}^t$. $(\cdot)^H$ is the Hermitian transpose.

Channel extrapolation has been proposed to extrapolate the complete channel matrix $\textbf{H}$ using a limited number of $\{h_{n_\text{R}n_\text{T}}\}$ \cite{zhang2023ai,gao2025enabling}, which is expressed mathematically as: 
\begin{equation}
\widehat{\textbf{H}}=f_\theta(\{h_{n_\text{R}n_\text{T}}\}),
\end{equation}where $\widehat{\textbf{H}}$ is the extrapolated channel matrix. $f_\theta(\{h_{n_\text{R}n_\text{T}}\})$ is the function that extrapolates the complete channel matrix using partial known CSI $\{h_{n_\text{R}n_\text{T}}\}$, which is acquired by training an AI model using massive channel data \cite{kim2020massive,chen2024channel,stenhammar2024comparison,jiang2019recurrent,peng2020lstm,zhang2022predicting,fan2024e2enet,jin2025linformer,SSnet2025gao,fan2025csi,lin2025generative,catak2025bert4mimo,sheng2025wireless,jiang2025mimo}. $f_\theta(\{h_{n_\text{R}n_\text{T}}\})$ is essential the correlation between $h_{n_\text{R}n_\text{T}}$ for different Tx-Rx pairs, which is learned generally by minimizing the mean squared error (MSE) between the extrapolated channel matrix $\widehat{\boldsymbol{H}}$ and the true channel matrix $\boldsymbol{H}$:
\begin{equation}
     \min_{f'_\theta}(|\widehat{\textbf{H}'}-\textbf{H}|^2).
\end{equation}

The generalization of existing AI-driven approaches is challenging, as the key factors of propagation characteristics, including the number of propagation paths $L_{n_\text{R}n_\text{T}}$, the complex amplitude $\alpha_{l_{n_\text{R}n_\text{T}}}$ and phase $\phi_{l_{n_\text{R}n_\text{T}}}$ of each path, the receiving array response $\mathbf{A}_\text{R}$ and transmuting array response $\mathbf{A}_\text{T}$, varying dramatically in different scenarios, thereby affecting the CSI correlation in antenna-domain. For example, the number of propagation paths $L_{n_\text{R}n_\text{T}}$ in indoor scenarios is generally much larger than that of the outdoor scenarios due to much large number of scatters in the indoor scenarios \cite{du202128,diago2020bringing,guo2025integrated,ju2024statistical}. The complex amplitude $\alpha_{l_{n_\text{R}n_\text{T}}}$ and phase $\phi_{l_{n_\text{R}n_\text{T}}}$ of each path are highly-related to the LoS/NLoS condition, carrier frequency, etc. The receiving array response $\mathbf{A}_\text{R}$ and transmitting array response $\mathbf{A}_\text{T}$ are highly-related to the location of BS and UE, and the physical environment. 

\begin{figure*}[htbp]
  \centering
  \includegraphics[width=0.83\linewidth]{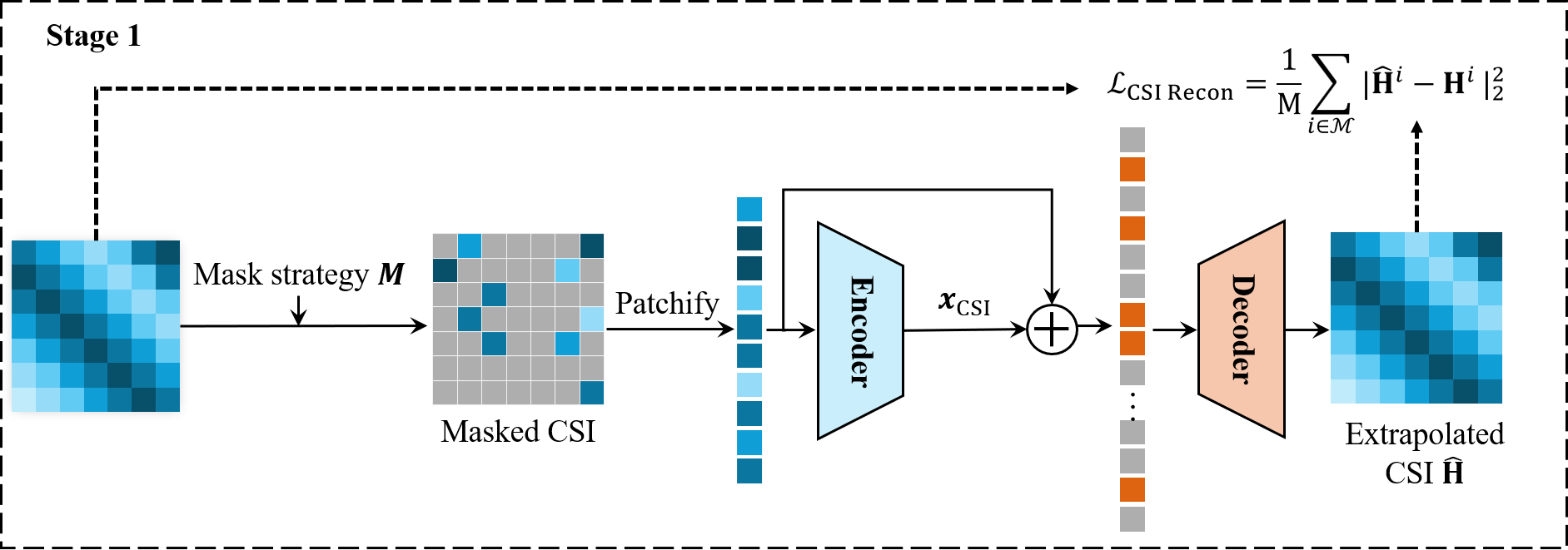} 
  \caption{Stage 1: experts emergent. Experts emergent by pre-training a masked auto-encoder (MAE)-based model via masked channel stata information (CSI) reconstruction. Inspired by \cite{xiao2024configurable,zhang2023emergent}, we expect the feed-forward network (FFN) layers to exhibit functional modularity after pre-training, which can later be exploited by clustering.}
  \label{fig:stage1}
\end{figure*}

\subsection*{Proposed architecture}
\label{Sec:3}
We next present our methodology for channel extrapolation as illustrated in Fig. \ref{fig:whole_arch}, which first pre-trains a model using MAE, then reorganizes the encoder FFN layers into specialized expert modules, and finally fine-tunes the adapted model separately across diverse propagation scenarios.

\subsection*{Experts Emergent via Pre-training}
As illustrated in Fig. \ref{fig:stage1}, we start by pre-training an MAE-based channel extrapolation model. As such MAE model has been widely adopted in wireless communications in \cite{zayat2023transformer,liu2025wifo} and our previous work in \cite{gao2026multiass}, we just briefly introduce the architecture of MAE. Interested readers can refer to the above references for implementation detail. 

The MAE contains an encoder and a decoder, where the function of the encoder is extracting the latent features $\mathbf{Z}_\text{CSI}\in\mathbb{R}^{B\times L\times D}$ of the masked CSI matrix $\mathbf{X}_\text{CSI}\in\mathbb{R}^{B \times 2 \times N_\text{T} \times N_\text{R}}$, where $B$, $L$ and $D$ denote the batch size, number of patches in a CSI matrix, and dimension of the positional embedding matrix, respectively. The masked CSI matrix $\mathbf{X}_\text{CSI}$ is obtained by randomly masking the complete CSI matrix $\textbf{H}$, i.e., $\mathbf{X}_\text{CSI}=\textbf{H}\circ\textbf{M}$, where $\textbf{M}\in \left\{0,1\right\}^{B \times M\times K}$ is the binary mask, equaling zeros for the masked locations and ones for the preserved positions. The ratio of the number of zeros in $\textbf{M}$ to $BMK$ is the mask ratio, which generally closes 1 as limited overhead are allocated for pilots to acquire CSI. $\textbf{A}\circ \textbf{B}$ denote the Hadamard product of the matrices $\textbf{A}$ and $\textbf{B}$. To preserve the low-level channel information contained in the encoder input, we introduce a residual connection. Specifically, the encoder input is directly added to the latent features:
\begin{equation}
    \widetilde{\mathbf{Z}}_{\text{CSI}}
    =
    \mathbf{Z}_{\text{CSI}}
    +
    \mathbf{X}_{\text{CSI}}.
    \label{eq:encoder_residual}
\end{equation}

The addition is performed element-wise because
$\mathbf{Z}_{\text{CSI}}$ and $\mathbf{X}_{\text{CSI}}$ have the same
token length and embedding dimension. The residual-enhanced
representation $\widetilde{\mathbf{Z}}_{\mathrm{CSI}}$ is then fed into
the decoder to reconstruct the complete CSI matrix:
\begin{equation}
    \widehat{\mathbf{H}}
    =
    \mathcal{D}
    \left(
        \widetilde{\mathbf{Z}}_{\text{CSI}}
    \right),
\end{equation}
where $\mathcal{D}(\cdot)$ denotes the decoder. This residual pathway
enables the decoder to exploit both the high-level representations
extracted by the encoder and the low-level channel information preserved
in the embedded input tokens.

The pre-training objective is to minimize the mean squared error (MSE) loss, expressed as:
\begin{equation}
\mathcal{L}_{\text{CSI Recon}} = \frac{1}{\mathcal{M} } \sum_{i \in \mathcal{M}}^{M} \left\| \textbf{H}_i-\widehat{\textbf{H}}_i \right\|
^2,
\end{equation} where $\mathcal{M}$ is the set of masked CSI indices for training dataset.

Generally, the FFN layers account for about two-thirds of Transformer's parameters, and achieve the following computation: 
\begin{equation}
\text{FFN}(\mathbf{X}) = \text{GELU}(\mathbf{X}\mathbf{W}_1 + \mathbf{b}_1)\mathbf{W}_2 + \mathbf{b}_2\label{FFN},    
\end{equation}where $\mathbf{W}_1$ and $\mathbf{W}_2$ are the learnable weights. $\mathbf{b}_1$ and $\mathbf{b}_2$ are the learnable biases. $\text{GELU}()$ is the GELU activation. As illustrated in \cite{xiao2024configurable,zhang2023emergent}, the FFN layers in Transformer-based model show function modularity after pre-training in NLP, i.e., a structure that groups neurons into modules by function, and each module works for its corresponding function. Inspired by the above research, we expect to obtain emerging experts for channel extrapolation across various scenarios in the FFN layers. 

\begin{figure}[tbp]
  \centering
  \includegraphics[width=1\linewidth]{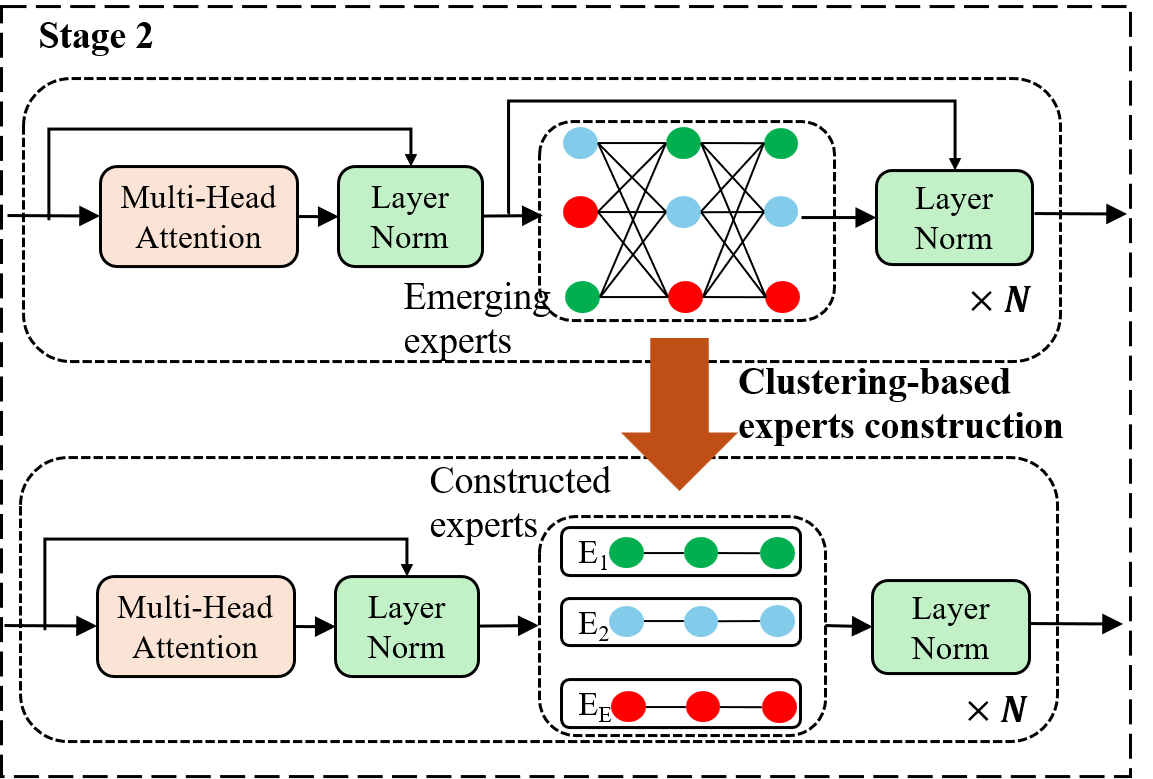} 
  \caption{Stage 2: experts construction. Experts constructed by weight-space clustering of the neurons in the  feed-forward network (FFN) of the encoders of the pre-trained masked auto-encoder (MAE) model. To facilitate balanced computation and efficient parallel execution, all experts are constructed with the same number of hidden neurons.}
  \label{fig:stage2}
\end{figure}
\begin{figure*}[htbp]
  \centering
  \includegraphics[width=0.83\linewidth]{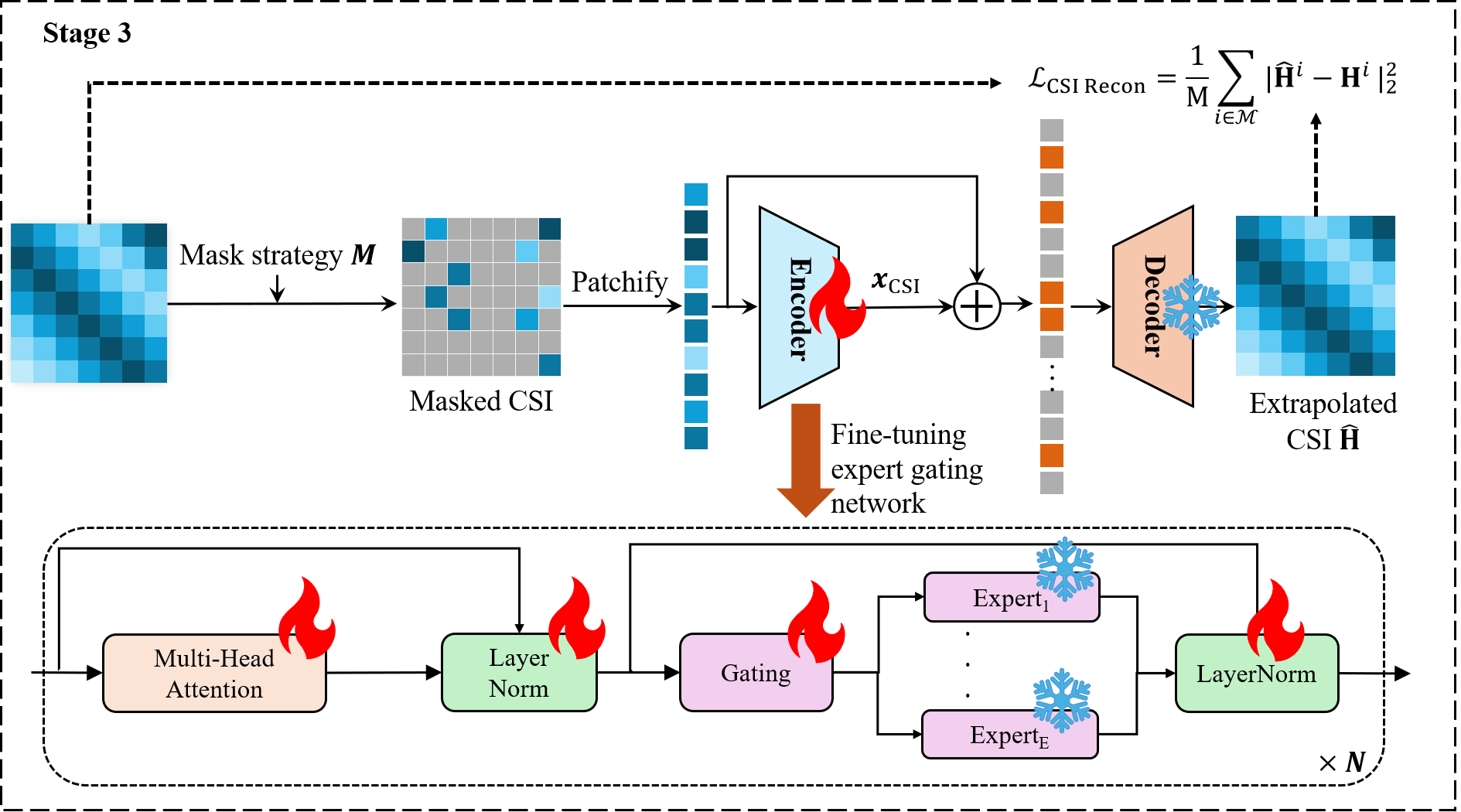} 
  \caption{Stage 3: experts selection. After completing the expert construction of the FFN layers in the pre-trained MAE encoder, a lightweight gating network is proposed to enable dynamic expert selection by fine-tuning the remaining model except the experts. Modules marked with `red fires' are fine-tuned. Modules marked with `snow flake' are kept frozen.}
  \label{fig:stage3}
\end{figure*}

\begin{figure*}
    \centering
    \subfigure[Indoor LoS.]{
        \includegraphics[width=0.45\linewidth]{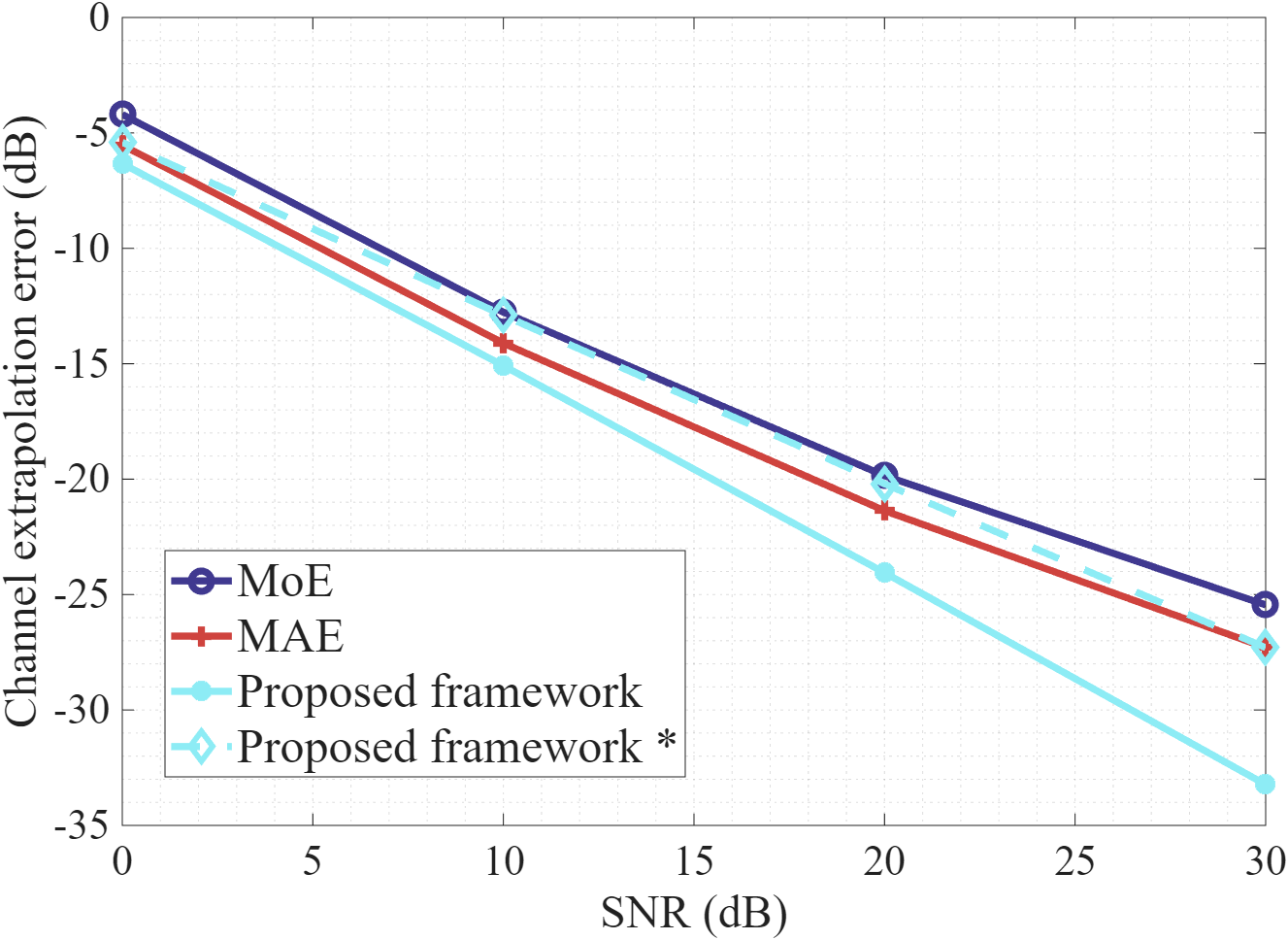}
        \label{fig:indis_LoSIn}
    }  
    \subfigure[Outdoor NLoS.]{
        \includegraphics[width=0.45\linewidth]{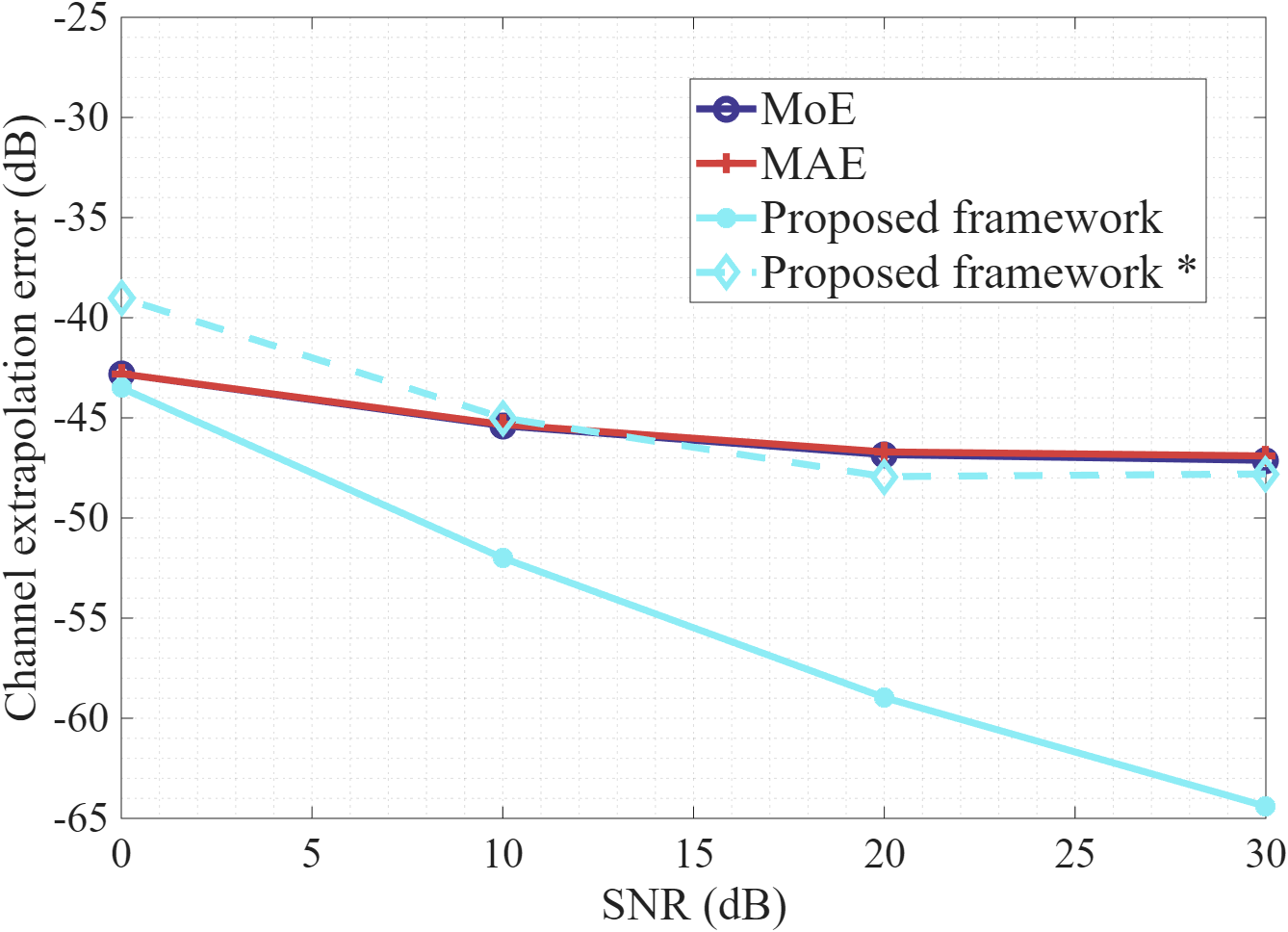}
        \label{fig:indis_NLoSOut}
    }  
    \caption{In-distribution performance. The in-distribution performance comparison between the mixed of expert MoE \cite{SSnet2025gao}, masked auto-encoder (MAE) \cite{he2022masked}, the proposed framework and the propose framework * with random expert selection. Figs \ref{fig:indis_LoSIn} and \ref{fig:indis_NLoSOut} illustrates the channel extrapolation performance in indoor line-of-sight (LoS) and outdoor none-LoS (NLoS) scenarios, respectively. The dark blue solid lines with circle markers represent the MoE baseline. The red solid lines with cross markers represent the MAE baseline. The cyan solid lines with diamond markers represent the proposed framework. The cyan dashed lines with diamond markers represent the proposed model with random experts selection (denoted as Proposed framework *). The proposed configurable framework consistently outperforms both the MoE and MAE baselines. The performance gain primarily arises from the three-stage process that emerges and clusters specialized experts and fine-tunes the gating network. An ablation study that retains the same experts but replaces the learned gating network with random expert selection yields substantially higher channel extrapolation error, confirming the critical role of the trained gating mechanism.}
    \label{fig:indis}
\end{figure*}

\subsection*{Expert Construction}

After pre-training, we adopt a parameter-clustering-based method
\cite{zhang2022moefication} to partition each FFN layer in the pre-trained
MAE encoder into multiple functionally specialized expert modules, as
illustrated in Fig.~\ref{fig:stage2}. The core idea is to group hidden
neurons with similar parameter representations into the same expert.
To facilitate balanced computation and efficient parallel execution, all
experts are constructed with the same number of hidden neurons.

Following the row-vector convention, an FFN layer is formulated as
\begin{equation}
    \operatorname{FFN}(\mathbf{x})
    =
    \operatorname{GELU}
    \left(
        \mathbf{x}\mathbf{W}_1 + \mathbf{b}_1
    \right)
    \mathbf{W}_2
    +
    \mathbf{b}_2,
    \label{FFN}
\end{equation}
where
$\mathbf{W}_1 \in \mathbb{R}^{d_{\mathrm{model}} \times d_{\mathrm{ff}}}$,
$\mathbf{b}_1 \in \mathbb{R}^{d_{\mathrm{ff}}}$,
$\mathbf{W}_2 \in \mathbb{R}^{d_{\mathrm{ff}} \times d_{\mathrm{model}}}$,
and
$\mathbf{b}_2 \in \mathbb{R}^{d_{\mathrm{model}}}$.
Under this formulation, the $n$-th hidden neuron is jointly associated
with the $n$-th column of $\mathbf{W}_1$, the $n$-th entry of
$\mathbf{b}_1$, and the $n$-th row of $\mathbf{W}_2$.

When partitioning the FFN layer into $k$ experts, each expert preserves
the input and output dimension $d_{\mathrm{model}}$, while its intermediate
dimension is reduced to
\begin{equation}
    d_e = \frac{d_{\mathrm{ff}}}{k}.
\end{equation}
The parameters of the $i$-th expert are therefore defined as
\begin{equation}
    \mathbf{W}_1^{(i)}
    \in
    \mathbb{R}^{d_{\mathrm{model}} \times d_e},
    \mathbf{b}_1^{(i)}
    \in
    \mathbb{R}^{d_e},
    \mathbf{W}_2^{(i)}
    \in
    \mathbb{R}^{d_e \times d_{\mathrm{model}}}.
\end{equation}

To construct the experts, we adopt a balanced clustering strategy based
on parameter similarity. Specifically, each column
$\mathbf{W}_1[:,n] \in \mathbb{R}^{d_{\mathrm{model}}}$ is treated as the
feature representation of the $n$-th hidden neuron. We then apply balanced
K-Means \cite{malinen2014balanced} to these neuron representations. The
algorithm partitions the $d_{\mathrm{ff}}$ hidden neurons into $k$
clusters, with each cluster containing exactly $d_e$ neurons.

Let $e:\{1,2,\ldots,d_{\mathrm{ff}}\}\rightarrow\{1,2,\ldots,k\} $ denote the neuron-to-expert assignment function. The index set associated
with the $i$-th expert is defined as:
\begin{equation}
    \mathcal{I}_i
    =
    \left\{
        n \mid e(n)=i
    \right\},
    |\mathcal{I}_i| = d_e.
\end{equation}

Balanced K-Means is performed through an iterative assignment-and-update
procedure. First, $k$ cluster centroids are initialized using the
K-Means strategy. Each neuron representation is then assigned to a
centroid while enforcing the equal-size constraint. Subsequently, each
centroid is updated as the mean of the neuron representations assigned
to the corresponding cluster. These steps are repeated to minimize
\begin{equation}
    \min_{\{\mathcal{I}_i,\boldsymbol{\mu}_i\}_{i=1}^{k}}
    \sum_{i=1}^{k}
    \sum_{n \in \mathcal{I}_i}
    \left\|
        \mathbf{W}_1[:,n] - \boldsymbol{\mu}_i
    \right\|_2^2,
    \qquad
    \text{s.t.}
    |\mathcal{I}_i| = d_e,
    \forall i.
\end{equation}

After obtaining the neuron-to-expert assignment, all parameters associated with the same hidden-neuron indices are reorganized consistently. For the $i$-th expert, the corresponding parameters are extracted as
\begin{equation}
    \mathbf{W}_1^{(i)}
    =
    \mathbf{W}_1[:,\mathcal{I}_i],
    \mathbf{b}_1^{(i)}
    =
    \mathbf{b}_1[\mathcal{I}_i],
    \mathbf{W}_2^{(i)}
    =
    \mathbf{W}_2[\mathcal{I}_i,:],
    \label{eq:expert_parameter_reorganization}
\end{equation}
where the same neuron index set $\mathcal{I}_i$ is used to select the
corresponding columns of $\mathbf{W}_1$, entries of $\mathbf{b}_1$, and
rows of $\mathbf{W}_2$. The output bias $\mathbf{b}_2$ is not partitioned
along the hidden-neuron dimension. Instead, it is retained as a shared
output bias and is added once after the selected expert outputs are
aggregated.

This indexing-based parameter reorganization preserves the original
hidden-neuron correspondence among $\mathbf{W}_1$, $\mathbf{b}_1$, and
$\mathbf{W}_2$. It is mathematically equivalent to first permuting the
hidden neurons according to their cluster assignments and then splitting
the permuted FFN parameters into equal-sized blocks, while avoiding the
explicit construction of permutation matrices.

\subsection*{Expert Selection}
After completing the expert construction of the FFN layers in the pre-trained MAE encoder, we introduce a lightweight gating network to enable dynamic expert selection. As illustrated in Fig. \ref{fig:stage3}, the network takes the input $\mathbf{X}_\text{CSI}$ and processes it through a gating to generate expert activation scores:
\begin{equation}
{g}_e = \text{softmax}\left(\text{gating}(\mathbf{X}_\text{CSI})\right), 
\label{eq:gating}
\end{equation}where $\text{softmax}()$ is the softmax function and the parameters in the gating function $\text{gating}()$ are fine-tuned in this stage.

We apply a top-$r$ selection strategy where only the experts with the $r$ highest scores are activated. The outputs of the selected experts are aggregated via a weighted sum based on their normalized scores, allowing the model to dynamically combine specialized transformations according to the input characteristics. This design maintains efficient computation by activating only a subset of experts while enabling the model to develop and utilize specialized expert functionalities effectively.
\subsection*{Remarks}
After the above three-stage framework, we obtain the same `MoE' model proposed in \cite{SSnet2025gao} in a fundamental different way from a modular perspective.

The MoE structure in \cite{SSnet2025gao} is pre-defined, especially the clustering of experts; while the experts in this manuscript are constructed by weight-space clustering of the corresponding columns in $\mathbf{W}_1$. For fair comparison, we adopt the same expert structure, i.e., the number of experts in each FFN layer, the number of neurons for each expert.

The gating function in \cite{SSnet2025gao} is trained along with the whole model. A load-balance policy is applied to ensure that all the experts are equally-well-trained, avoiding the weights and biases of some experts are over-updated, while those of other experts are under-updated. Nevertheless, the gating function is trained in the third stage of the proposed framework, given the construction of experts (in stage 2) after emergent via pre-training in the first stage. We aim the select the experts contributing most to the channel extrapolation in a specific scenario without applying load-balance policy. For fair comparison, we adopt the same gating model for both \cite{SSnet2025gao} and the proposed framework.

Although we apply the MAE as our backbone, the proposed three-stage architecture is applicable to any Transformer-based model for various tasks in wireless communication, such as channel extrapolation, resource allocation, precoding, etc., across scenarios.  
%\subsection{Fine-tuning}

\begin{figure*}[t]
    \centering
    \subfigure[Indoor NLoS.]{
        \includegraphics[width=0.45\linewidth]{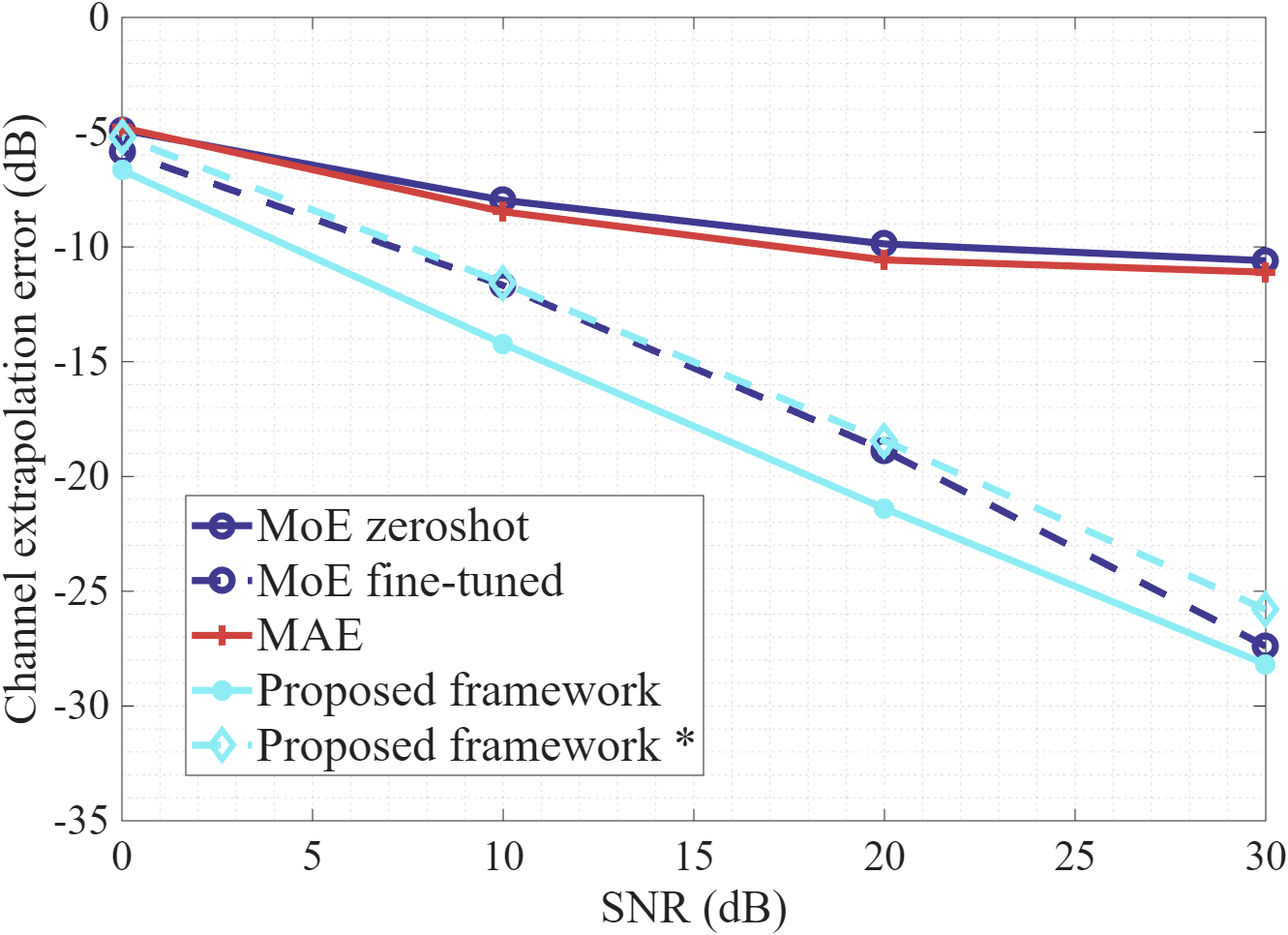}
        \label{fig:outdis_NLoSIn}
    } 
    \subfigure[Outdoor LoS.]{
        \includegraphics[width=0.45\linewidth]{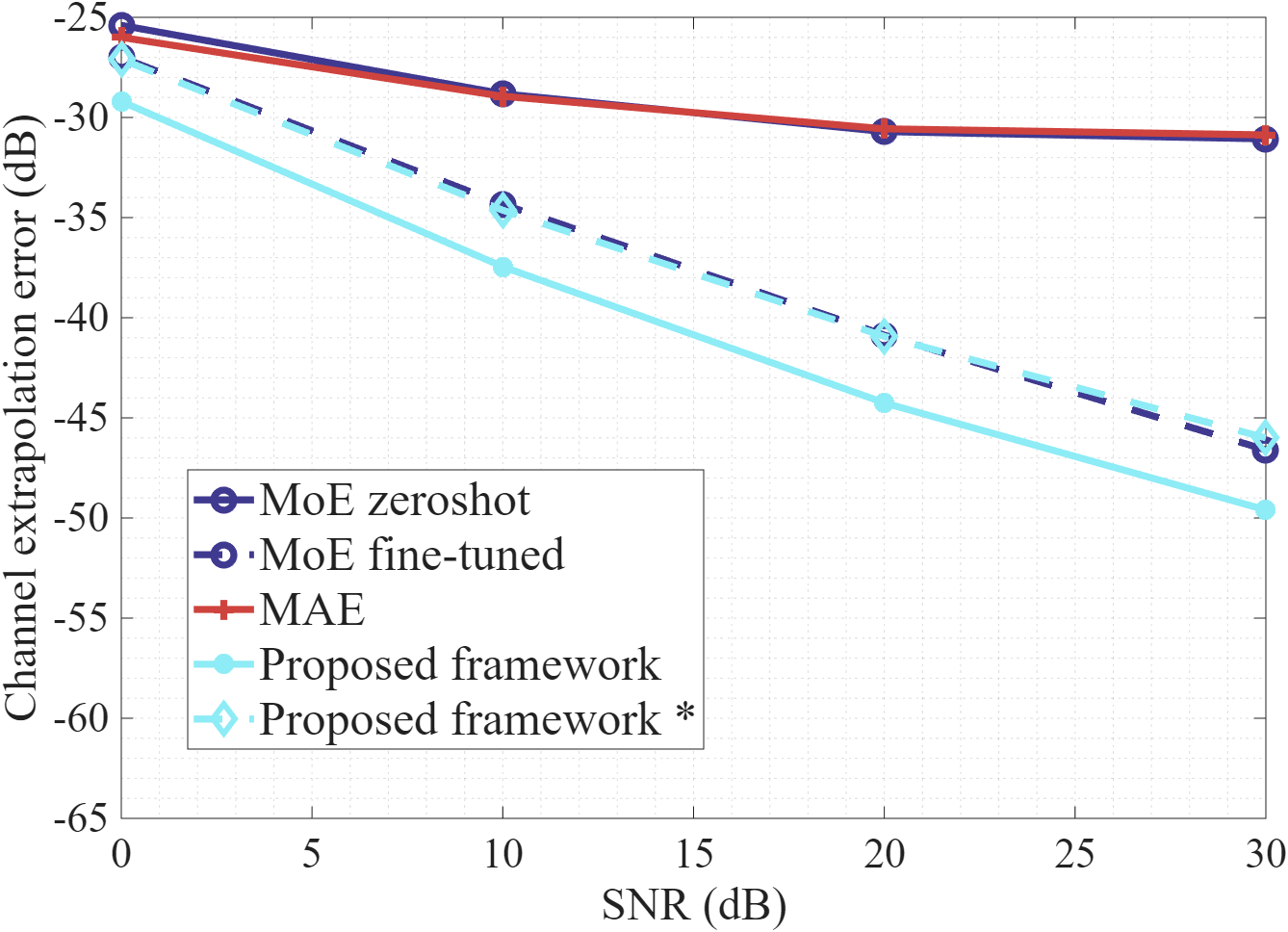}
        \label{fig:outdis_LoSOut}
    }  
    \caption{Out-of-distribution performance. The out-of-distribution performance comparison between the zeroshot and fine-tuned mixed of expert (MoE) \cite{SSnet2025gao}, masked auto-encoder (MAE) \cite{he2022masked}, the proposed framework and the propose framework * with random expert selection. Figs \ref{fig:outdis_NLoSIn} and \ref{fig:outdis_LoSOut} illustrates the channel extrapolation performance in indoor none-line-of-sight (NLoS) and outdoor line-of-sight (LoS) scenarios, respectively. The dark blue solid lines with circle markers represent the MoE model evaluated in the zero-shot setting. The dark blue dashed lines with circle markers represent the MoE model after fine-tuning. The red solid lines with cross markers represent the MAE baseline. The cyan solid lines with diamond markers represent the proposed framework. The cyan dashed lines with diamond markers represent the proposed model with random experts selection (denoted as Proposed framework *). Even when the MoE baseline is further fine-tuned on the target domains, the proposed framework still achieves clearly superior performance and demonstrates stronger generalization. The same random-expert-selection ablation again produces markedly higher error across the entire SNR range, further verifying that the learned gating network, rather than the experts alone, is essential for the observed robustness.}
    \label{fig:outdis}
\end{figure*}
\begin{figure*}
    \centering
    \subfigure[Indoor LoS.]{
    \includegraphics[width=0.45\linewidth]{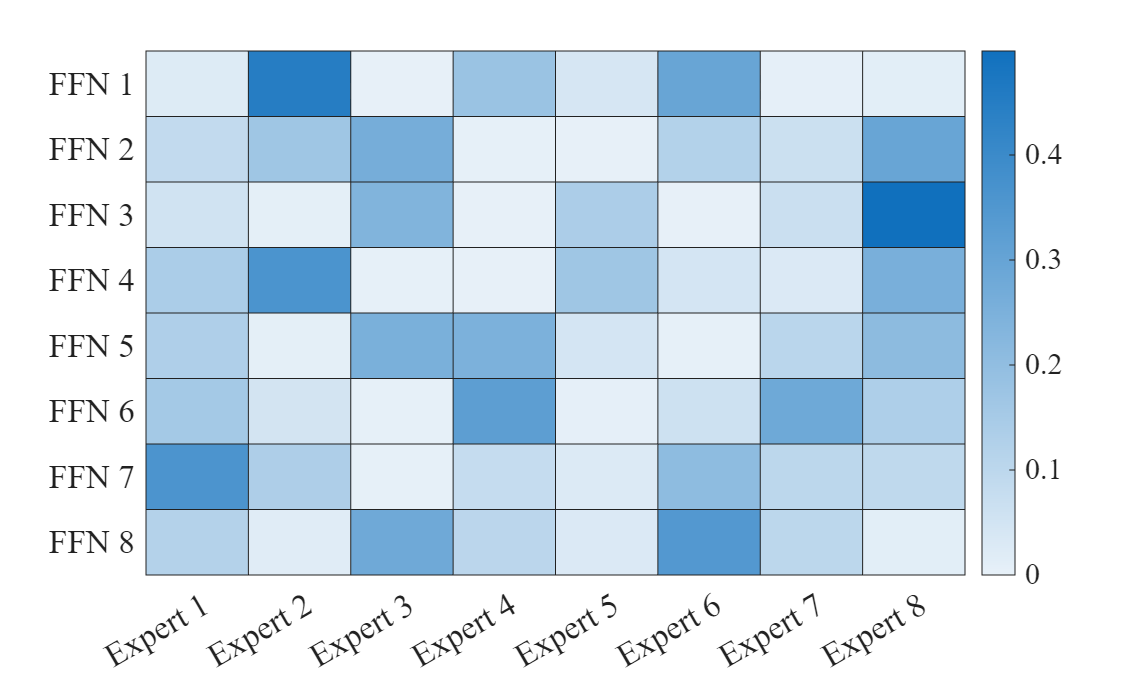}
        \label{fig:heatmap_LoS_indoor}
    }  
    \subfigure[Outdoor NLoS.]{
    \includegraphics[width=0.45\linewidth]{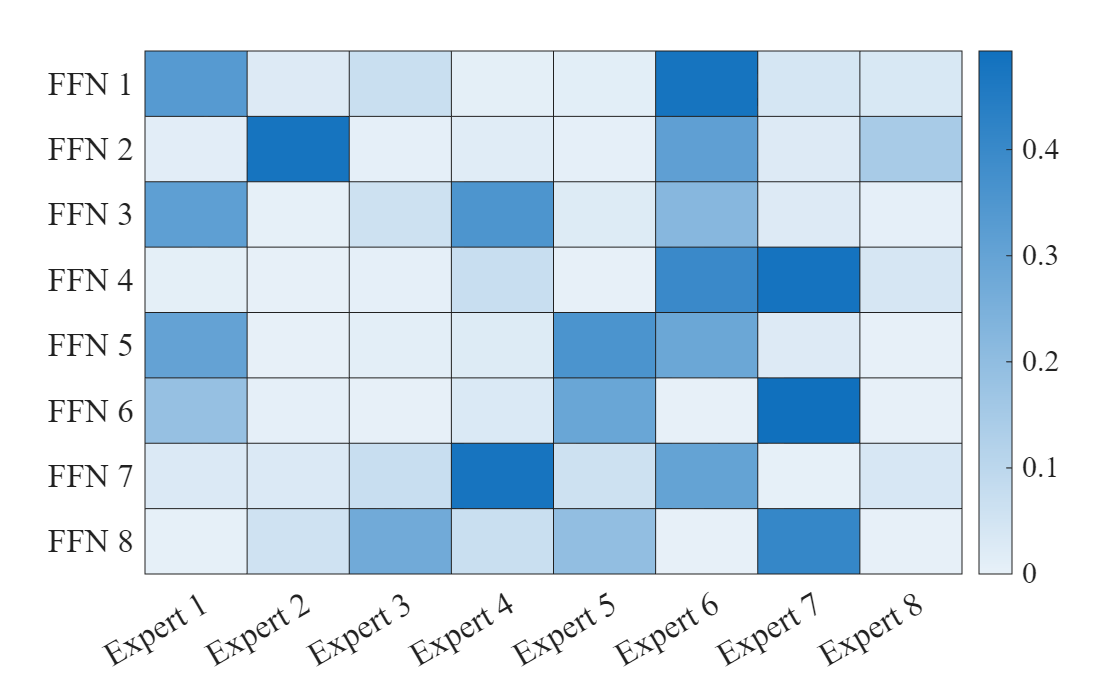}
        \label{fig:heatmap_NLoS_outdoor}
    } 
            \subfigure[Indoor NLoS.]{
    \includegraphics[width=0.45\linewidth]{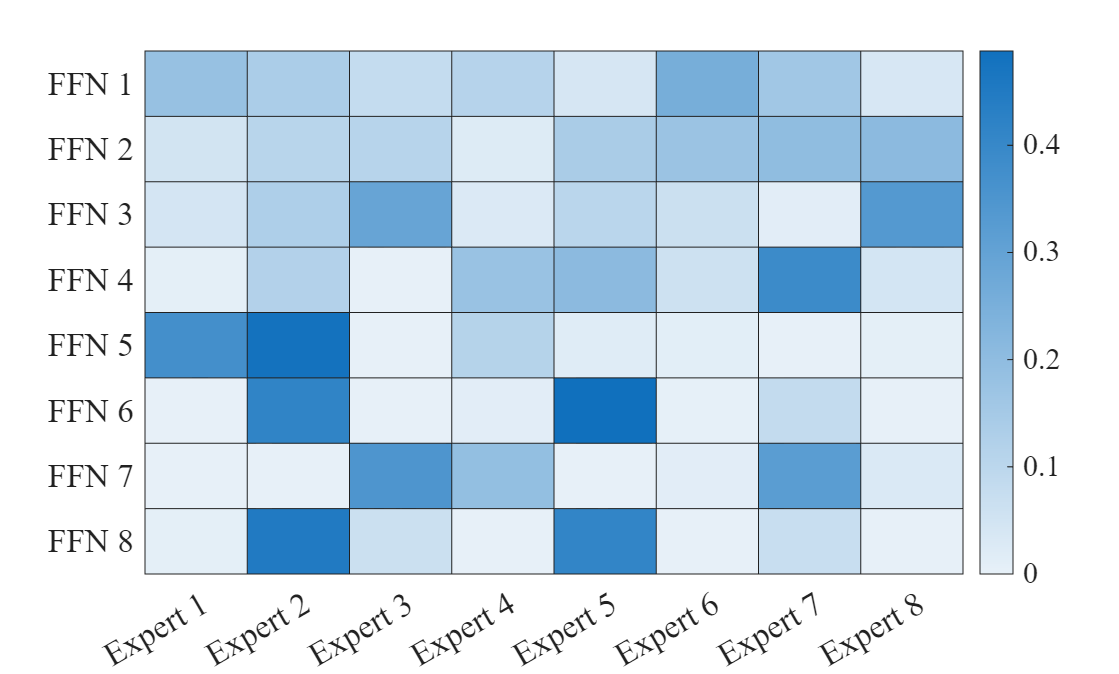}
        \label{fig:heatmap_NLoS_indoor}
    } 
        \subfigure[Outdoor LoS.]{
    \includegraphics[width=0.45\linewidth]{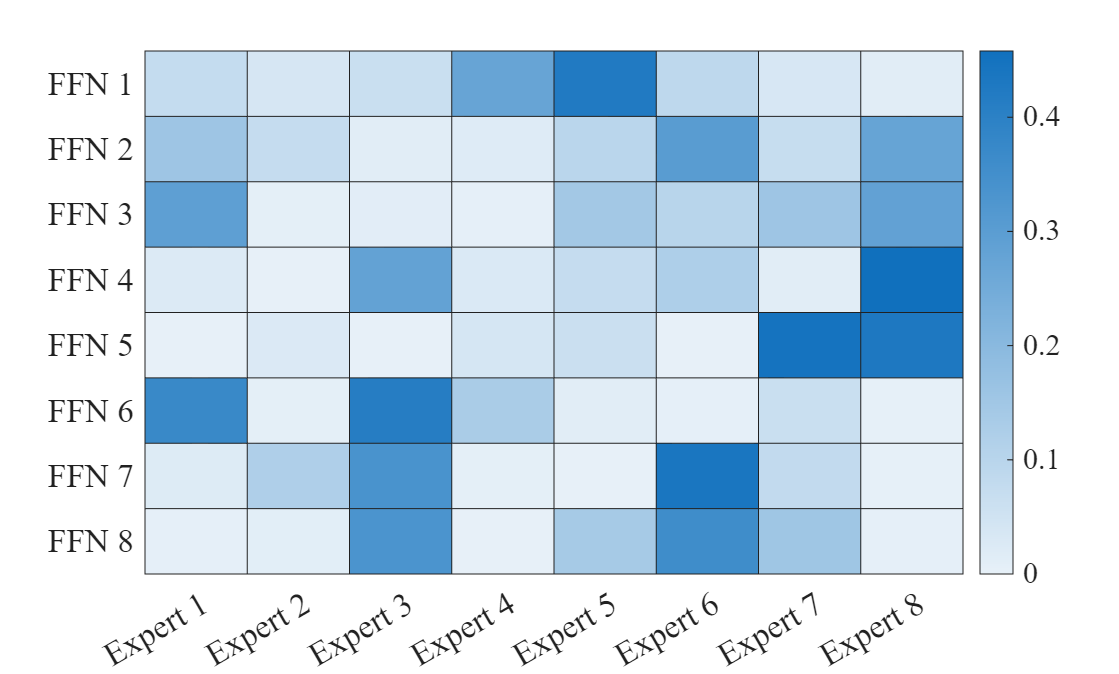}
        \label{fig:heatmap_LoS_outdoor}
    } 
    \caption{Experts activation probabilities in each scenario. The illustration of the activation probability of experts in the proposed model for indoor line-of-sight (LoS), outdoor none-LoS (NLoS), indoor NLoS and outdoor LoS in Figs \ref{fig:heatmap_LoS_indoor}, \ref{fig:heatmap_NLoS_outdoor}, \ref{fig:heatmap_NLoS_indoor} and \ref{fig:heatmap_LoS_outdoor}, respectively. Obvious sparsification of the activation probability of the proposed framework are observed in different scenarios.}
    \label{fig:expert_dis}
\end{figure*}
\section*{Results}
\subsection*{Data generation}
In our experiments, we use DeepMIMO \cite{alkhateeb2019deepmimo}, a ray-tracing-based MIMO generator, and QuaDRiGa \cite{6758357,Jaeckel2023QuaDRiGa}, a 3GPP-compliant channel simulator, to generate channel data. The center frequencies of both the DeepMIMO and QuaDRiGa datasets are set to 60 GHz. We select the `O1' and `I3' scenarios in DeepMIMO. The system bandwidth is set to 160 MHz with 200 OFDM subcarriers, resulting in a subcarrier spacing of 0.8 MHz. For the outdoor scenario `O1', we activate BS2 and BS3. User Grid 1 is activated to generate outdoor LoS data, while User Grid 2 and User Grid 3 are activated to generate outdoor NLoS data. For the indoor scenario `I3', we activate BS1 and BS2. The designated LoS User Grid is activated to generate indoor LoS data, and the NLoS User Grid is activated to generate indoor NLoS data. To further evaluate the generalization capability of the proposed framework, we additionally generate industrial indoor factory (InF) datasets using QuaDRiGa, where half of the datasets are configured under the \texttt{3GPP\_38.901\_InF\_LOS} condition, while the remaining datasets follow the \texttt{3GPP\_38.901\_InF\_NLOS\_DH} condition. Specifically, the QuaDRiGa simulation uses a single BS located at a height of 3 m and 20,000 stationary UEs uniformly distributed over a $100~\mathrm{m}\times100~\mathrm{m}$ area at a height of 1.5 m. The BS is equipped with an $8\times2$ uniform planar array with 16 antennas, while each UE employs an $8\times1$ uniform linear array with 8 antennas. Both arrays use vertically polarized 3GPP millimeter-wave antenna elements with half-wavelength spacing and a $7^\circ$ electrical downtilt. The system bandwidth is set to 100 MHz with 256 OFDM subcarriers, resulting in a subcarrier spacing of 0.39 MHz. InF is a scenario with rich scatterers, making its channel characteristics substantially different from those of the indoor and outdoor scenarios. In total, we collect 80,000 CSI samples, including 20,000 samples for the indoor LoS scenario and 20,000 samples for the outdoor NLoS scenario, as well as 10,000 samples for each of the indoor NLoS, outdoor LoS, InF LoS, and InF NLoS scenarios. Each CSI sample can be denoted as $\mathbf{h}_m\in\mathbb{C}^{16\times8}$, where 16 and 8 refer to the numbers of transmit and receive antennas, respectively.

\subsection*{Training settings}
The model was implemented using the PyTorch framework and trained on an NVIDIA GeForce RTX 4090 GPU. For the baseline model MAE, we combine the 50\% indoor LoS data and 50\% outdoor NLoS data to form training data. Based on this, we divide the encoder FFN layer into $K$ experts. Then, we fixed this layer parameters and fine-tune in each scenario. In the experts selection stage, we adopt a two-layer network to generate expert activation scores and apply a top-$r$ selection strategy. Specifically, the first linear layer projects the input to a hidden dimension, followed by a $\text{tanh}$ activation function that introduces non-linearity while maintaining gradient stability. The second linear layer then maps these activated features to output scores corresponding to each available expert. For fair comparison, the proposed framework and the baseline MOE employ the identical number of experts per FFN layer, the same  intermediate dimension per expert, the same top-r routing budget, and the same gating architecture. The only structural difference is the origin of the experts themselves: the MoE of \cite{SSnet2025gao} learns its expert parameters and router jointly from scratch (or from a random initialization of the same architecture), whereas the proposed method freezes the experts obtained by balanced K-means clustering of the pre-trained.

Unless otherwise specified, we set $K=8$ and $r=2$ across all experiments. To analyze the impact of expert numbers, we conduct an ablation study by varying $(K,r)\in{(4,1),(8,2),(16,3)}$, evaluating the trade-off between representation ability and computational overhead.

\section*{Discussion}
%\subsection*{Channel Extrapolation Performance}

\begin{figure}
  \centering
  \includegraphics[width=1\linewidth]{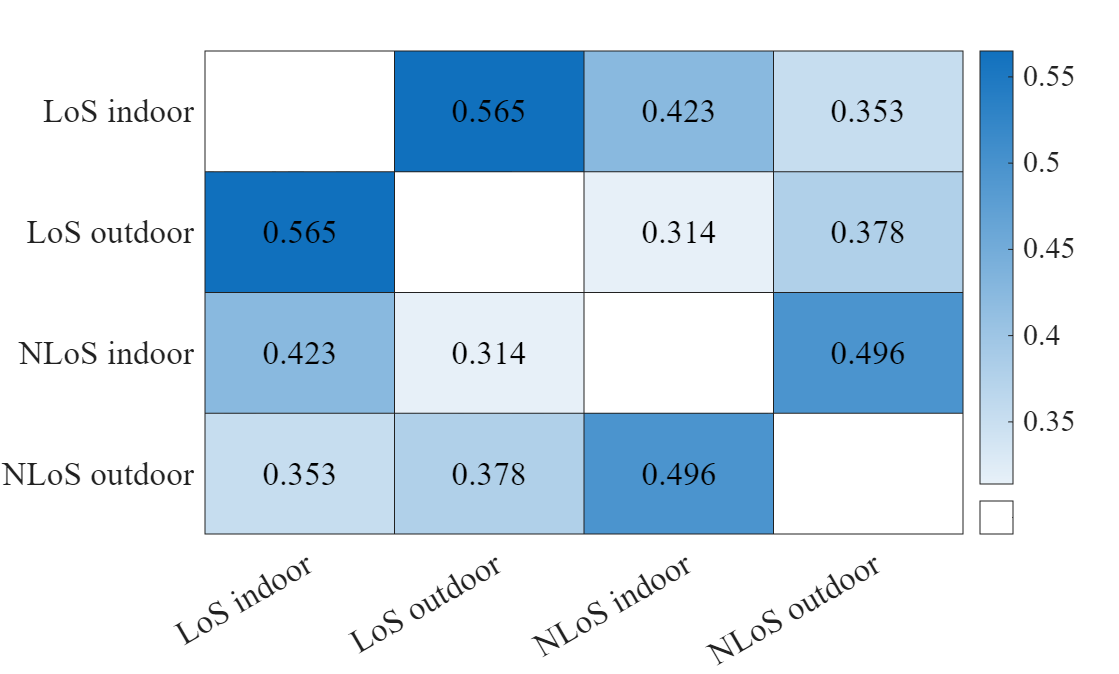} 
  \caption{Experts activation patterns between scenarios. The cosine similarity of the experts activation probability in the proposed model for indoor line-of-sight (LoS), outdoor none-LoS (NLoS), indoor NLoS and outdoor LoS. The similarity of expert activation probabilities across different scenarios shows that relatively similar environments produce more similar expert usage patterns. Specifically, the highest similarity occurs between indoor LoS and outdoor LoS, followed by indoor NLoS and outdoor NLoS, while the lowest similarities appear between indoor LoS and outdoor NLoS (and vice versa).}
  \label{fig:experts_similarity}
\end{figure}
\begin{figure*}
    \centering
    \subfigure[Indoor LoS.]{
    \includegraphics[width=0.45\linewidth]{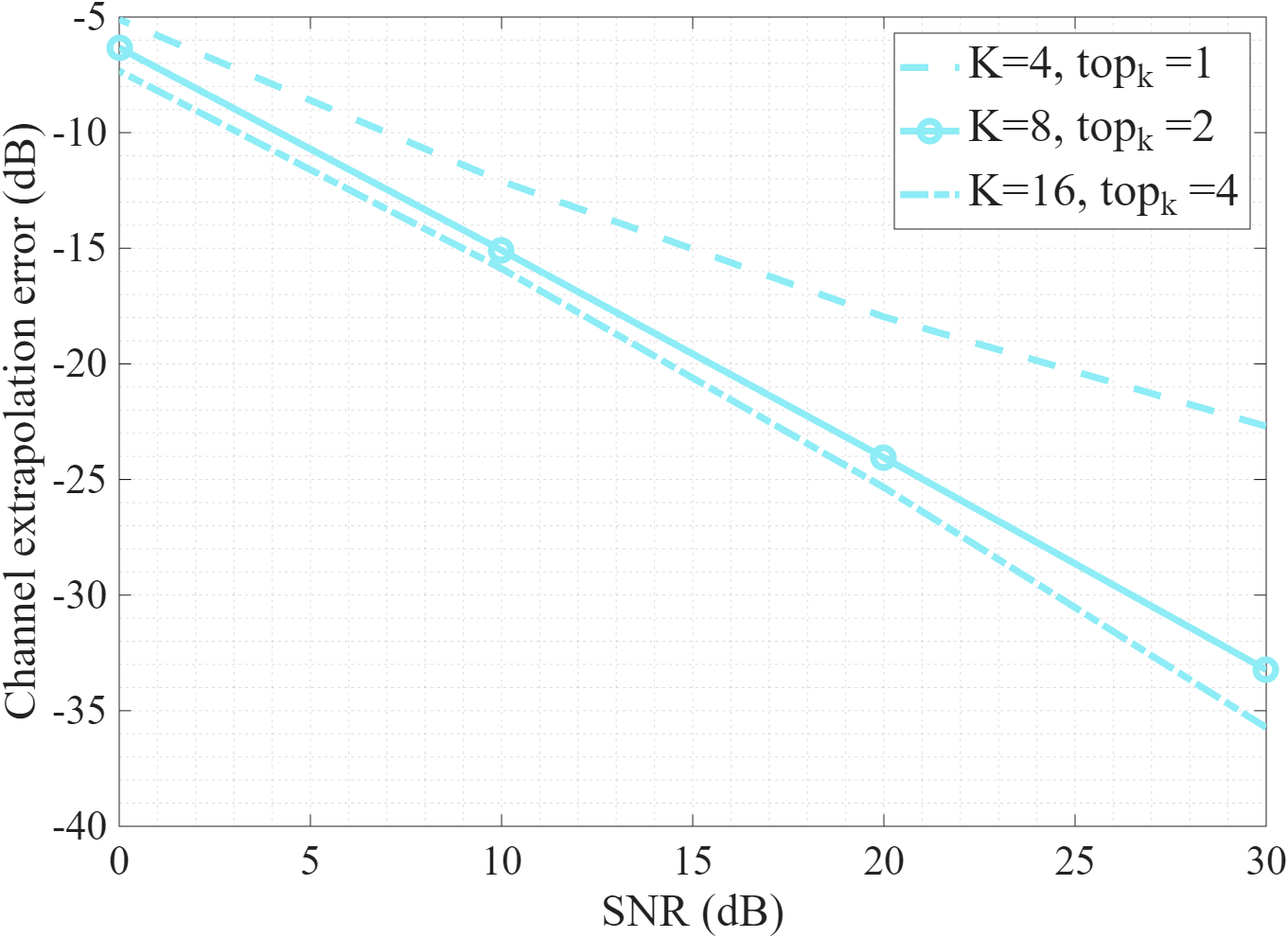}
        \label{fig:indis_LoSIn_varying_K}
    }  
    \subfigure[Outdoor NLoS.]{
    \includegraphics[width=0.45\linewidth]{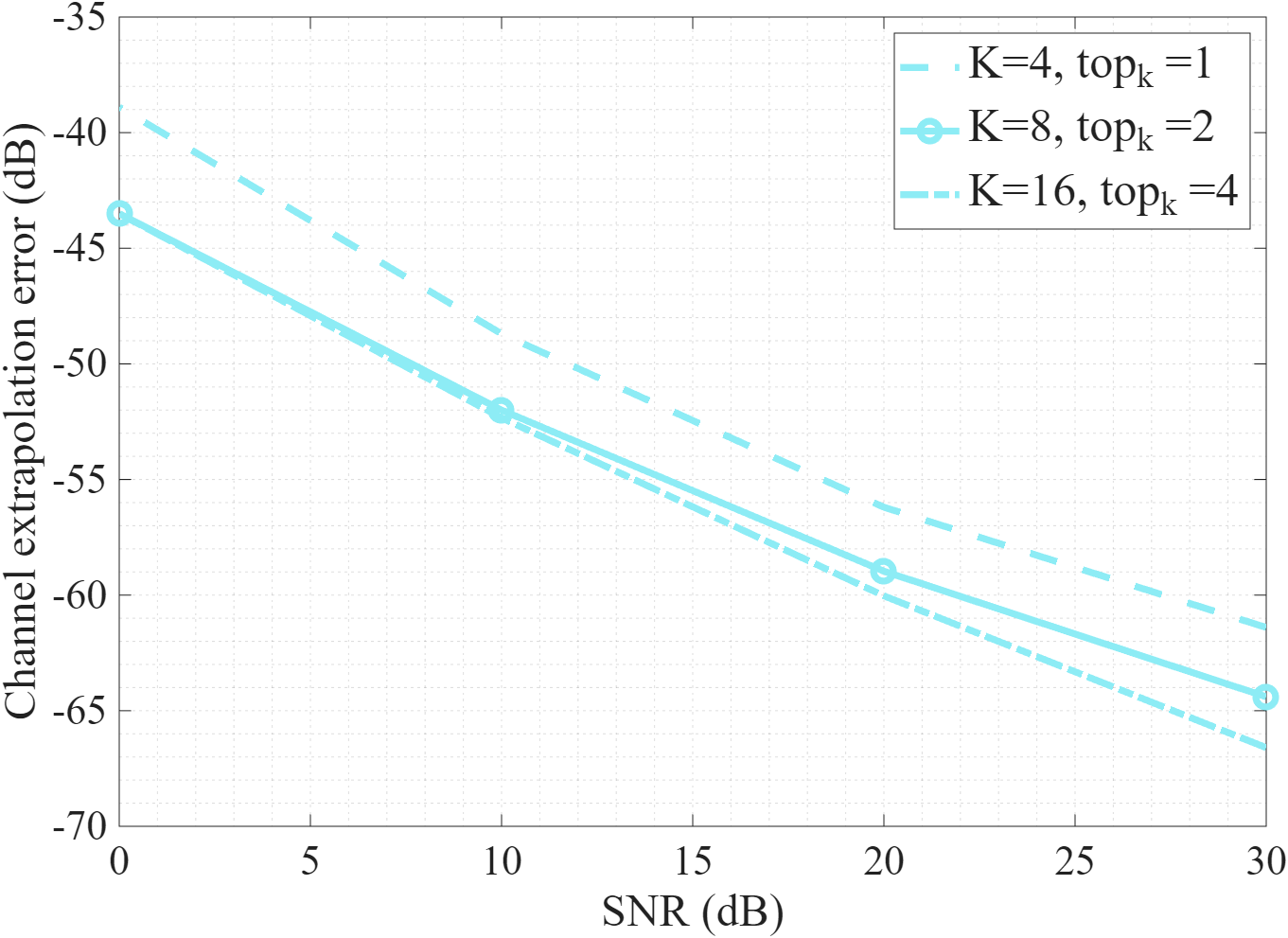}
        \label{fig:indis_NLoSOut_varying_K}
    } 
            \subfigure[Indoor NLoS.]{
    \includegraphics[width=0.45\linewidth]{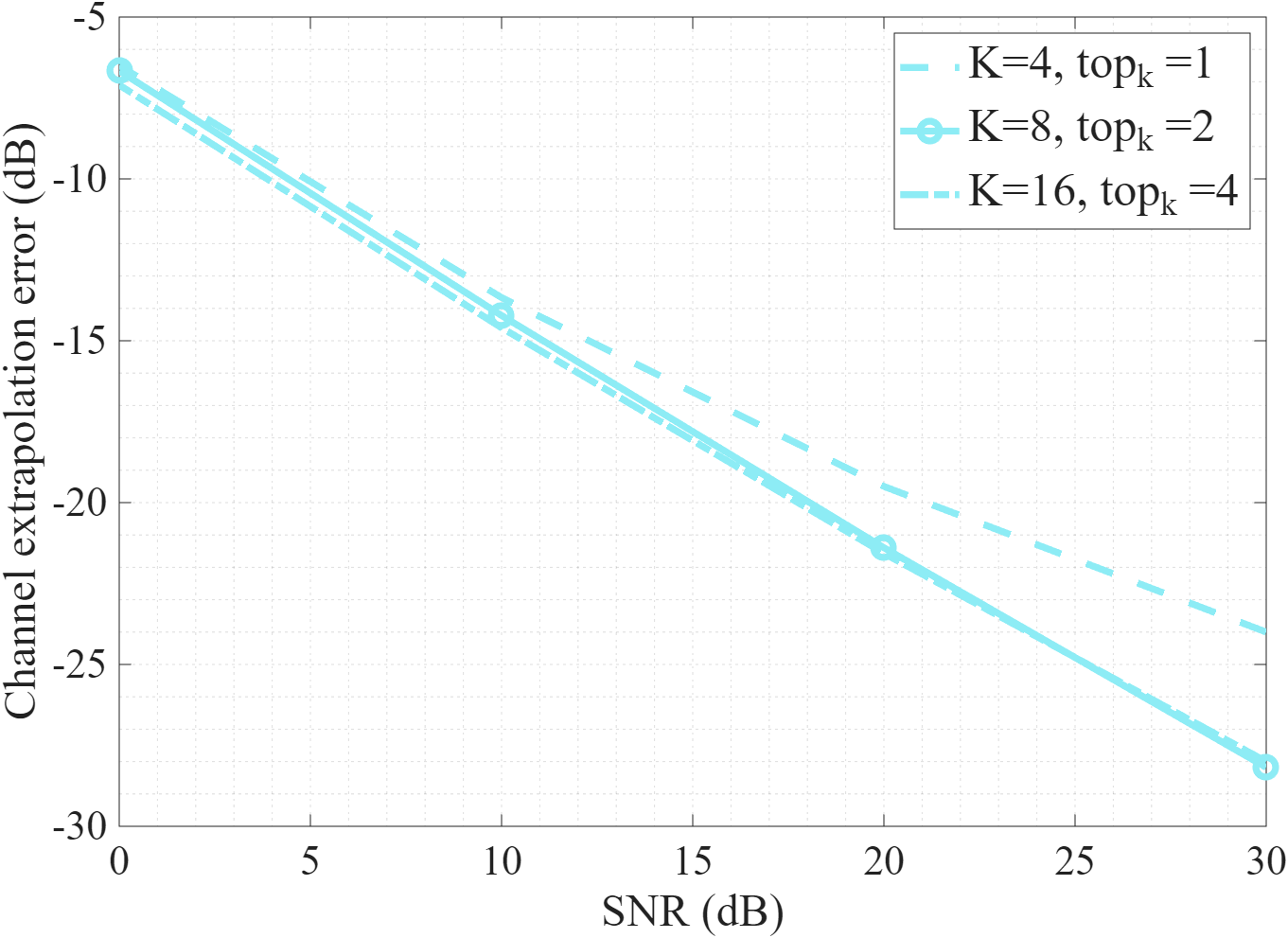}
        \label{fig:outdis_NLoSIn_varying_K}
    } 
        \subfigure[Outdoor LoS.]{
    \includegraphics[width=0.45\linewidth]{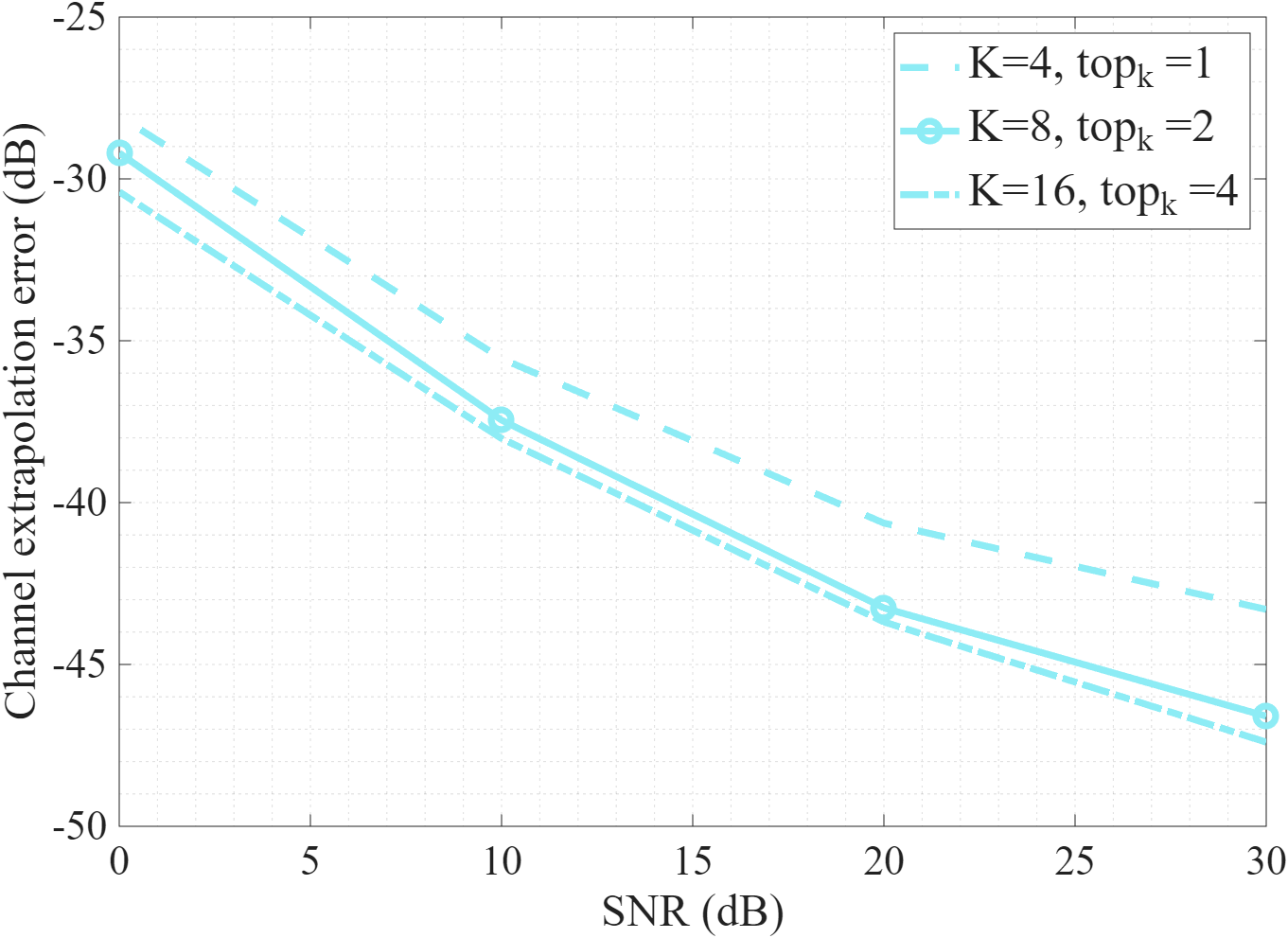}
        \label{fig:outdis_LoSOut_varying_K}
    } 
    \caption{Performance with respect to the number of experts. Channel extrapolation performance of the proposed framework with the number of experts at a fixed activation ratio (i.e., $1/4$) for indoor line-of-sight (LoS), outdoor none-LoS (NLoS), indoor NLoS and outdoor LoS in Figs \ref{fig:indis_LoSIn_varying_K}, \ref{fig:indis_NLoSOut_varying_K}, \ref{fig:outdis_NLoSIn_varying_K} and \ref{fig:outdis_LoSOut_varying_K}, respectively. The dotted cyan lines without markers represent the configuration with $\mathrm{K}=4$ and $\mathrm{top}_k=\mathrm{1}$. The solid cyan lines with circle markers represent the configuration with $\mathrm{K}=8$ and $\mathrm{top}_\mathrm{k}=2$. The dashed cyan lines without markers represent the configuration with $\mathrm{K}=16$ and $\mathrm{top}_k=\mathrm{4}$. Increasing the number of experts from 4 to 8 yields a clear reduction in channel extrapolation error across all scenarios, whereas a further increase to 16 brings only marginal additional gains. This consistent ranking in both in-distribution and out-of-distribution settings indicates that a moderate number of experts is already sufficient to capture the dominant modular structure of chanenl state information (CSI) correlations.}
    \label{fig:varying_K}
\end{figure*}
\begin{figure*}
    \centering
    \subfigure[Indoor factory LoS.]{
    \includegraphics[width=0.48\linewidth]{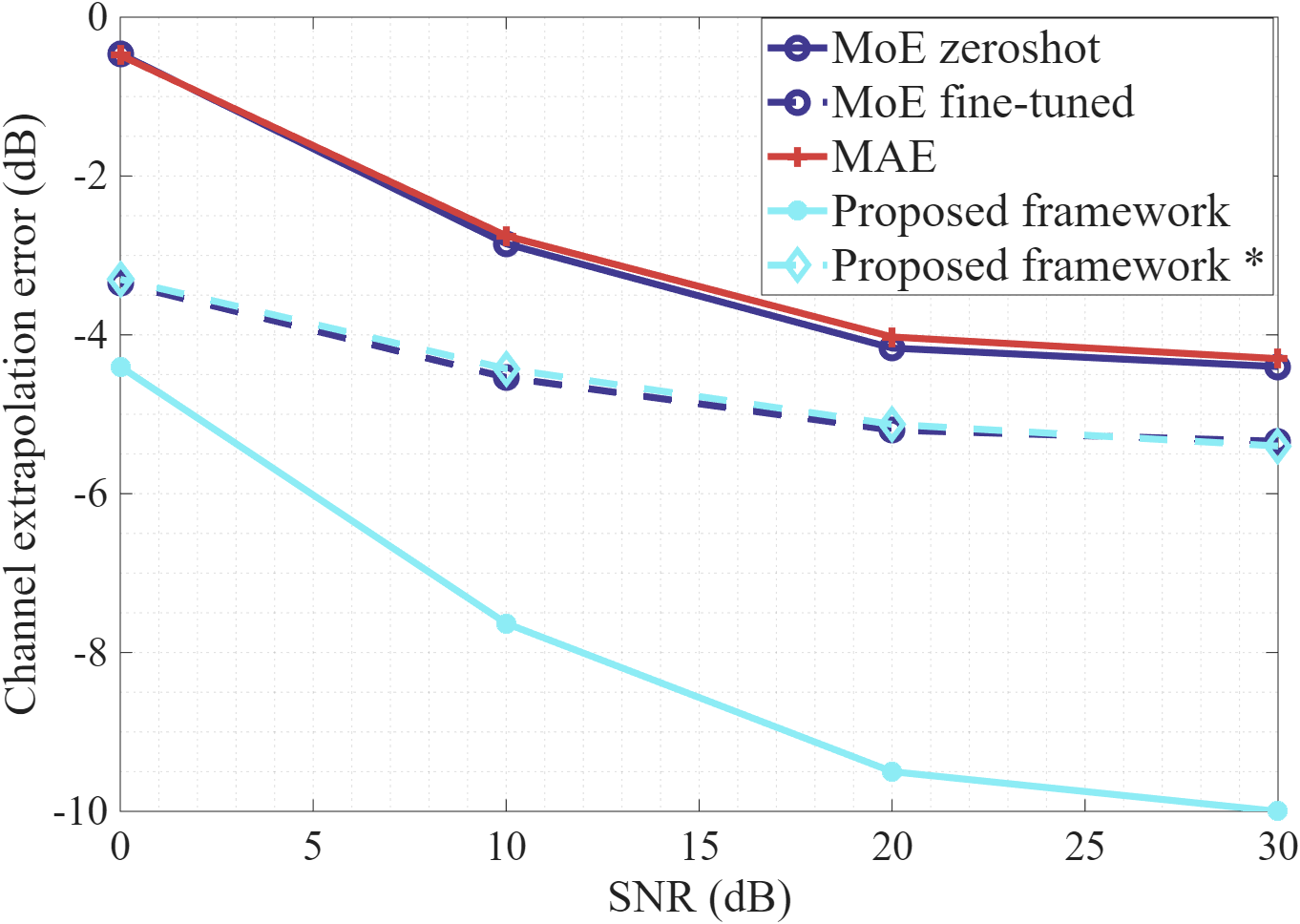}
        \label{fig:LoSInf}
    }  
    \subfigure[Indoor factory NLoS.]{
    \includegraphics[width=0.48\linewidth]{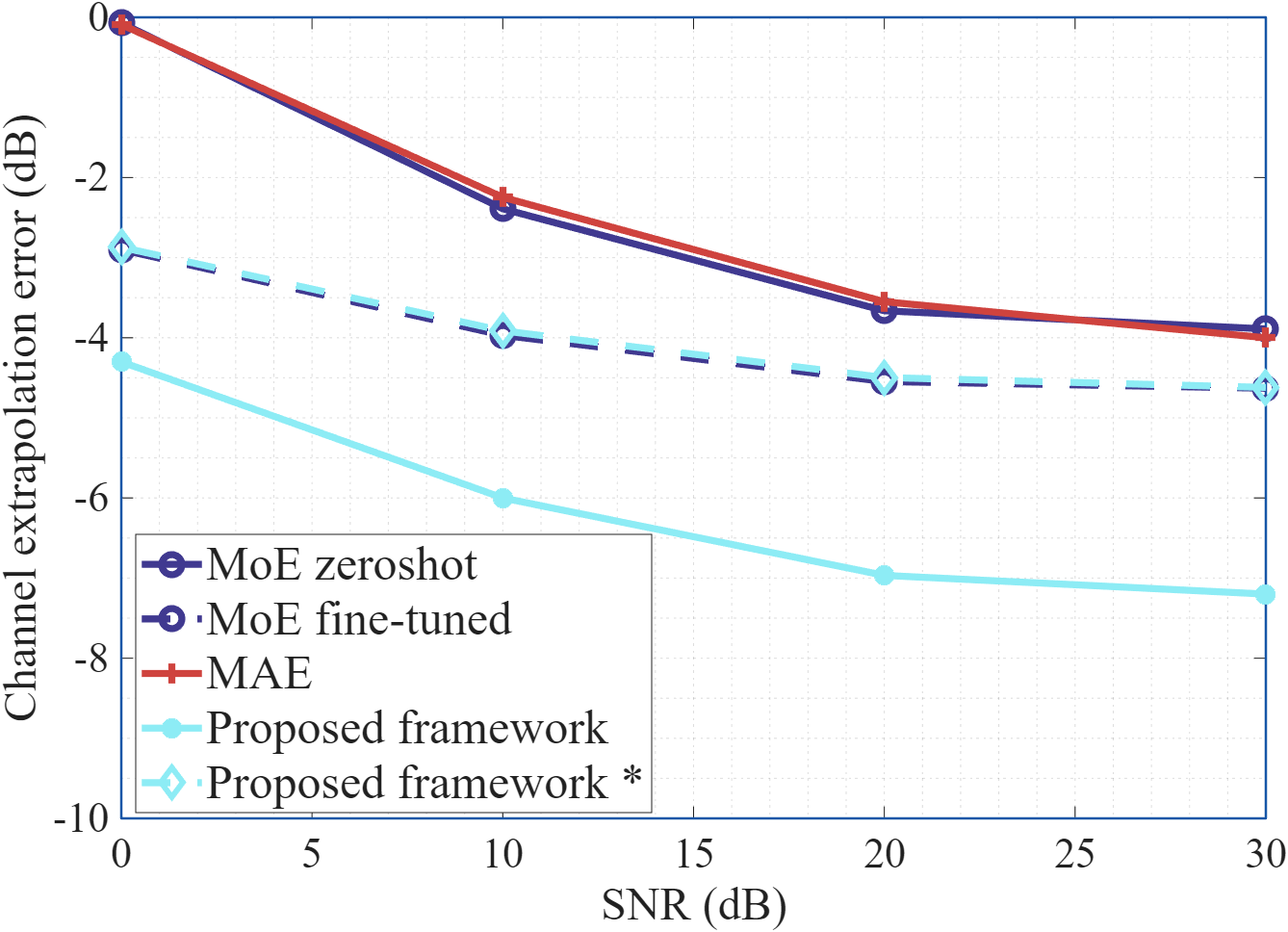}
        \label{fig:NLoSInf}
    }  
    \caption{Performance in indoor factory (InF) scenarios. The performance comparison between the mixed of expert (MoE) \cite{SSnet2025gao}, masked auto-encoder (MAE) \cite{he2022masked}, the proposed framework and the propose framework * with random expert selection for InF generated using QuaDRiGa channel simulator. The results of InF line-of-sight (LoS) and InF none-LoS (NLoS) are illustrated in Figs. \ref{fig:LoSInf} \ref{fig:NLoSInf}, respectively. The dark blue solid lines with circle markers represent the MoE model evaluated in the zero-shot setting. The dark blue dashed lines with circle markers represent the MoE model after fine-tuning. The red solid lines with cross markers represent the MAE baseline. The cyan solid lines with diamond markers represent the proposed framework. The cyan dashed lines with diamond markers represent the proposed model with random experts selection (denoted as Proposed framework *). the proposed framework achieves the lowest channel extrapolation error on both InF LoS and InF NLoS channels, outperforming zero-shot MoE, fine-tuned MoE, MAE, and the random-expert-selection variant across the entire SNR range. These results demonstrate that the experts learned in Stage 2 capture transferable modular structure that remains effective even in a previously unseen indoor-factory environment.}
    \label{fig:inf}
\end{figure*}

The proposed configurable framework is compared with the MoE \cite{SSnet2025gao}, MAE \cite{he2022masked} in both the in-distribution and out-of-distribution manner.
%\subsubsection*{In-distribution comparison}
Fig. \ref{fig:indis} illustrates the performance of the proposed configurable framework and the baseline models in-distribution manner, i.e., the training and testing datasets are simulated in the same scenarios, namely indoor LoS and outdoor NLoS scenarios. Figs. \ref{fig:indis_LoSIn} and \ref{fig:indis_NLoSOut} demonstrate that the proposed configurable framework outperforms the baseline models in indoor LoS and outdoor NLoS scenarios, respectively. In either indoor LoS or outdoor NLoS scenarios, MoE \cite{SSnet2025gao} outperforms MAE \cite{he2022masked} slightly, while the proposed configurable framework outperforms MoE \cite{SSnet2025gao} dramatically at relatively high SNR (greater than 10 dB). The differences between the proposed configurable framework and MoE \cite{SSnet2025gao} are the experts constitution and the experts gating network. By performing the proposed 3-stage framework, to emerge and cluster the experts and fine-tune the gating network, the channel extrapolation performance can be significantly enhanced.

Fig. \ref{fig:outdis} illustrate the performance of the proposed configurable framework and the baseline models in-distribution manner, i.e., the training datasets are simulated in the indoor LoS and outdoor NLoS scenarios, while the testing datasets are simulated in the indoor NLoS and outdoor LoS scenarios. Figs. \ref{fig:outdis_NLoSIn} and \ref{fig:outdis_LoSOut} demonstrate that the proposed configurable framework outperforms the baseline models in indoor NLoS and outdoor LoS scenarios, respectively. Different from the in-distribution comparison, the MoE \cite{SSnet2025gao} are further fine-tuned in either indoor NLoS or outdoor LoS scenarios and obtain a performance enhancement. However, the proposed configurable framework still outperform the fine-tuned MoE dramatically, not the mentioned the baseline MoE and MAE. In addition, the proposed framework shows higher generalization performance. Specifically, the proposed framework outperforms the best counterpart in out-of-distribution manner, i.e., fine-tuned MoE, by $1.3-16.2$ dB in indoor NLoS scenario, and by $1.7-18.3$ dB in outdoor LoS scenario. While the performance advantages of the proposed framework outperforms the best counterpart in in-distribution manner drop to $0.8-6.4$ in indoor LoS scenario, and $1.1-16.1$ dB in outdoor NLoS scenario, respectively. This indicates that the proposed framework shows superior generalization over baseline MoE \cite{SSnet2025gao}, MAE \cite{he2022masked}.

%\subsection*{Emerging Experts Analysis}
To further validate that the superior performance of the proposed framework attributed to the experts emergent and clustering, and the gating network for expert selection, we carry out ablation experiments for the proposed framework with random expert selection, in which the same experts are retained but the gating network is replaced by uniform random activation of 2 experts (denoted as Proposed framework *). Figs. \ref{fig:indis} and \ref{fig:outdis} show that this random variant produces markedly higher channel extrapolation error than the proposed framework in both ID and OOD scenarios. The performance gap remains large across the entire SNR range and is especially pronounced at moderate-to-high SNR.

These results confirm that the performance gains of the proposed framework do not arise merely from the existence of multiple experts. Proper routing that selects the experts most relevant to the input CSI is indispensable; without it the modular structure constructed in Stage 2 cannot be effectively exploited.

We further show the activation probability of the experts of the proposed configurable framework in Fig. \ref{fig:expert_dis}. Obvious sparsification of the activation probability of the proposed framework are observed in indoor LoS, outdoor NLoS, indoor NLoS and outdoor LoS in Figs. \ref{fig:heatmap_LoS_indoor}, \ref{fig:heatmap_NLoS_outdoor}, \ref{fig:heatmap_NLoS_indoor} and \ref{fig:heatmap_LoS_outdoor}, respectively. 

To further demonstrate that the emergent experts are highly-related to the channel propagation scenarios, the similarity of the experts probabilities in various scenarios is illustrated in Fig. \ref{fig:experts_similarity}. Intuitively, the similarity of the experts probabilities between relatively-similar scenarios is larger than that between relatively-different scenarios. This echoes with the results in Fig. \ref{fig:experts_similarity}, where the similarity between indoor LoS and outdoor LoS is the highest, then comes the indoor NLoS and outdoor NLoS, indicating that the LoS and NLoS affect the signal propagation more significantly than the indoor and outdoor scenarios. The similarities between indoor LoS and indoor NLoS, outdoor LoS and outdoor NLoS come next. The smallest similarities are between indoor LoS and outdoor NLoS, indoor NLoS and outdoor LoS. The results of our channel extrapolation analysis indicate that the LoS versus NLoS condition plays a more dominant role in determining channel characteristics than the distinction between indoor and outdoor environments. This is evidenced by the observed higher correlation in key propagation parameters, such as path loss, angular spread, and delay spread, between indoor LoS and outdoor LoS scenarios compared to that between indoor LoS and indoor NLoS scenarios. These findings align with established stochastic channel models, where LoS/NLoS classification serves as the primary determinant of large-scale and small-scale parameters, while indoor/outdoor distinctions primarily influence scenario-specific adjustments rather than core statistical behavior.

%\subsection*{Effectiveness of Experts Number}

The number of experts $  K  $ is a central modeling choice in the proposed framework. Because the method partitions the pre-trained FFN neurons into specialized modules, the value of $  K  $ directly controls the granularity of the resulting modular prior. An insufficiently large $  K  $ leaves each expert mixing multiple distinct CSI correlation patterns and thereby weakens modularization; an excessively large $  K  $ fragments the FFN, complicates router training, and risks eroding the function learned by the original MAE. 

Fig. \ref{fig:varying_K} illustrates the channel extrapolation performance of the proposed framework under different numbers of experts while the activation ratio is held fixed at 1/4. The configurations $  (K,r)=(4,1)  $, $  (8,2)  $ and $  (16,4)  $ are evaluated in the indoor LoS, outdoor NLoS, indoor NLoS and outdoor LoS scenarios. As shown in Figs. \ref{fig:indis_LoSIn_varying_K}-\ref{fig:outdis_LoSOut_varying_K} raising the number of experts from $4$ to $8$ yields a clear reduction in extrapolation error in every scenario. Increasing the number further to $16$ produces additional but smaller gains. These results indicate that a moderate number of experts is already sufficient to capture the dominant modular structure of CSI correlations that emerges during pre-training, while finer partitioning under the same activation budget brings diminishing returns. The same ranking of the three curves appears in both in-distribution and out-of-distribution settings, confirming that the expert-construction stage extracts modules that remain effective across propagation environments.

%\subsection*{Performance in different scenarios}
To further examine the generalization capability of the proposed framework, we generate InF channels with QuaDRiGa under the 3GPP TR 38.901 InF-DH model. The InF environment contains dense metallic scatterers and therefore exhibits multipath richness and angular spreads that differ markedly from the DeepMIMO indoor and outdoor scenarios used for pre-training and expert construction. Evaluating on this unseen domain provides a stringent test of whether the modular experts learned from the original data remain effective under substantially altered propagation conditions.

Fig. \ref{fig:inf} compares the proposed framework against MoE (zero-shot and fine-tuned), MAE, and a random-expert-selection variant on both InF LoS and InF NLoS channels. The proposed framework consistently yields the lowest extrapolation error across the entire SNR range. Relative to the best competing method (fine-tuned MoE), the gain reaches several dB at moderate-to-high SNR, while the random-selection baseline remains markedly inferior. These results confirm that the experts constructed in Stage 2 capture transferable modular structure rather than scenario-specific artifacts, enabling robust performance even in a propagation environment that was never observed during pre-training or fine-tuning.

%\subsection*{Computational Complexity Analysis}
\begin{table}[h]
\centering
\captionsetup{justification=justified, singlelinecheck=false}
\caption{Computational complexity comparison between the the MoE \cite{SSnet2025gao}, MAE \cite{he2022masked} and the proposed framework in terms of the No. of parameters for the entire model (FFN) and the FLOPs. The value in the $(*)$ indicates the ratio of the corresponding value to that of the proposed model. $\uparrow$, $\downarrow$ and $\leftarrow$ indicate increasing, decreasing and leveling, respectively. The value in \textbf{bold} indicates the best performance for each type of performance indicator. }
\begin{tabular}{l|c|c}
\hline
{Method}  & No. of parameters ($M$) & FLOPs  ($M$) \\
\hline
MoE \cite{SSnet2025gao} &${7.21} (4.22)$, $\leftarrow$ & $\textbf{27.61}$, $\leftarrow$\\
MAE \cite{he2022masked}  &$\textbf{7.16}$, $\downarrow7 \%$ & $38.22$, $\uparrow38 \%$\\
Proposed model  &${7.21} (4.22)$  & $\textbf{27.61}$ \\
\hline
\end{tabular}
\label{tab:inference_time}
\end{table}
Apart from generalization, we further evaluate the computational complexity-related issues, i.e., number of parameters for the entire model and the FFN layers ($M$), and FLOPs ($M$) of the proposed configurable framework in Table. \ref{tab:inference_time}. The proposed configurable framework has the same number of parameters as the MoE \cite{SSnet2025gao} (FFN layers account for approximately $60 \%$ of total number of parameters), which shares the exactly the structure, apart from the expert construction. Compared with the MAE \cite{he2022masked}, the proposed framework increases the number of parameters by $7$ \% by adding the expert selection gating network. However, despite a slight increase of the number of parameters, the FLOPs of the proposed framework is dramatically lower by 38 \% than that of the MAE \cite{he2022masked}, by sparsely activating the experts (FFN layers in the MAE).

\label{Sec:conclusions}

\ifCLASSOPTIONcaptionsoff
  \newpage
\fi
 \small
%\nocite{*}
\bibliographystyle{IEEEtran}
\bibliography{reference.bib}
\section*{Acknowledgment}
This work was supported by Shanghai Natural Science Foundation under Grant 25ZR1402148.
\section*{Author contribution}
Y.G. and S.X. conceived the idea and supervised the research. Y.G. designed the modular MAE framework and implemented the model. X.W. curated channel dataset, carried out model training and testing. X.W., J.J. and Y.Y. analyzed simulation results and prepared for Figs. 1-11 and Table. I. Y.G. wrote the original draft, assisted by X.W., J.J. and Y.Y. Y.J. and S.Z. provided expertise in the channel simulator and performance evaluation. Y.J., S.Z and Z.H. provided critical revisions to the manuscript and visualization of results. S.X. acquired funding, and reviewed and edited the final manuscript. All authors read and approved the submitted version.
\section*{Data availability}
All relevant data and figures supporting the main conclusions of the document are available on request. Please refer to Yuan Gao at gaoyuansie@shu.edu.cn.

\section*{Conflicts of Interest}
The authors declare no conflicts of interest.

\end{document}